\documentclass[12pt,nofootinbib,notitlepage,onecolumn,preprint,aps,
  pra,
  amsmath,amssymb,longbibliography]{revtex4-2}
\usepackage{ulem}
\usepackage[varg]{txfonts}
\usepackage{bm}
\usepackage{float}
\usepackage{setspace}
\usepackage{xcolor}
\usepackage[left]{lineno}
\usepackage[caption=false]{subfig}
\PassOptionsToPackage{hyphens}{url}
\usepackage[bookmarks=true,         
     pdfnewwindow=true,      
     colorlinks=true,    
     linkcolor=blue,     
     citecolor=blue,     
     filecolor=blue,  
     urlcolor=blue,      
     final=true,
 ]{hyperref}
\usepackage{color,epsfig,amsmath,amssymb,fancyhdr} 

\usepackage{footnote}
\newcommand\footnoteref[1]{\protected@xdef\@thefnmark{\ref{#1}}\@footnotemark}
\makeatother
\usepackage{etoolbox}
\usepackage{footmisc}
\makeatletter
\patchcmd{\frontmatter@abstract@produce}
  {\vskip200\p@\@plus1fil
   \penalty-200\relax
   \vskip-200\p@\@plus-1fil}
  {}
  {}
  {}
\makeatother

\usepackage{lipsum}

\usepackage{booktabs}

\begin{document}
\singlespacing

\title{Shear flows and their suppression at large aspect ratio. Two-dimensional simulations of a growing convection zone}

\author{J. R. Fuentes}
\author{A. Cumming}

\affiliation{Department of Physics and McGill Space Institute, McGill University, 3600 rue University, Montreal, QC H3A 2T8, Canada}

\begin{abstract}
We investigate the onset and evolution of zonal flows in a growing convective layer when a stably-stratified fluid with a composition gradient is cooled from above. 
This configuration allows the study of zonal flows for a wide range of values of the Rayleigh number, $Ra$, and aspect ratio of the convection zone within a given simulation. 
We perform a series of 2D simulations using the Boussinesq approximation, with aspect ratio of the computational domain between $1$ and $5$, and Prandtl number $Pr = 0.1$, 0.5, 1, and $7$.  For simulations with aspect ratio of one we find that zonal flows appear when the aspect ratio of the convective layer is smaller than two, and the evolution of the system depends on the Prandtl number.
For $Pr\leq 1$, the fluid experiences bursts of convective transport with negligible convective transport between bursts. The magnitude and frequency of the bursts are smaller at low $Pr$, which suggests that the bursting regime is stronger in a narrow range around $Pr=1$, as observed in previous studies of thermal convection. For $Pr=7$, the structure of the flow consists of tilted convective plumes, and the convective transport is sustained at all times. In wider domains, the aspect ratio of the convective zone is always much larger than two and zonal flows do not appear. These results confirm and extend to fluids with stable composition gradients previous findings on thermal convection. The fact that zonal flows can be avoided by using computational domains with large aspect ratios opens up the possibility of 2D studies of convective overshoot, layer formation and transport properties across diffusive interfaces.

\end{abstract}

\maketitle
\section{Introduction}

Large scale horizontal flows (hereafter zonal flows) arise in many geophysical and astrophysical fluids. Some examples include zonal jets in the ocean \citep{2006GeoRL..33.3605R}, the atmospheric super-rotation of Venus \citep{1970JAtS...27.1107T}, and zonal jets in giant planets \citep{1983PApGe.121..375B}. Although those systems are different in nature, their fluid motions often undergo convection and strong rotation. These features are relevant since their zonal flows might be a consequence of the interaction of convection and rotation \citep{1994Chaos...4..123B}. The mechanism of convection driven zonal flows can be understood as follows: shear perturbations resulting from rotation or any other horizontal anisotropy tilt convective motions, generating Reynolds stresses that reinforce the shear perturbations and therefore the amplitude of the horizontal flows. For an illustration, see Fig. 1 in \citet{1970JAtS...27.1107T}, or Fig. 2 in \citet{1983PApGe.121..375B}. This mechanism is called tilting instability and energy is supplied directly from the convective scales to the zonal flow \citep[see, e.g.,~][]{1993PhFlB...5..415F,1994PhPl....1..222R}.

In three dimensions, zonal flows have been observed only when anisotropies in the horizontal direction (such as rotation with the rotation axis misaligned with the gravity axis) are present \citep[e.g.,~][]{1992PhFlA...4.1333M,2020JFM...905A..21W}. An example is Rayleigh-Benard convection with rapid rotation around a horizontal axis \citep[see Fig. 1b in][for an illustration]{2020JFM...905A..21W}. Although early experiments of convection in water reported the formation of `winds' and large scale horizontal flows \citep{1954RSPSA.225..185M,Krishnamurti1981}, they are explained by convective rolls of large horizontal extent rather than resulting from the tilting instability \citep{PhysRevLett.91.064501}. Further, zonal flows have not been found in later laboratory experiments nor in three-dimensional numerical simulations of isotropic Rayleigh-Benard convection \citep{PhysRevLett.91.064501,2017PhRvF...2h3501A,2020JFM...905A..21W}.

Since small shear perturbations can form spontaneously in 2D-simulations of thermal convection, they are often used to study zonal flows and their properties at a much lower computational cost than 3D-simulations with rotation. In particular, 2D thermal convection with free-slip and periodic boundary conditions at large Rayleigh number\footnote{\label{note1}See Sect. \ref{sect_model} for a definition.} ($Ra$) has been used extensively as the canonical model to study zonal flows driven by convection. As pointed out by \citet{goluskin_johnston_flierl_spiegel_2014}, 1) two dimensionality prevents transverse perturbations that can reduce horizontal fluid motions, 2) periodic boundary conditions on the side boundaries do not confine the fluid in the horizontal direction, and 3) free-slip top and bottom boundaries apply no shear stresses to slow down the horizontal flow.

The main result from previous studies is that the nature of the flow and the transport properties depend strongly on the Prandtl number\footref{note1} ($Pr$) \citep{1993PhFlB...5..415F,PhysRevE.90.013017,2014PhFl...26e4104F,goluskin_johnston_flierl_spiegel_2014,PhysRevFluids.6.033502}. In particular, it has been shown that once the convective flow is affected by large-scale horizontal motions at large $Ra$, for low $Pr$ ($\lesssim 1$) it undergoes strong oscillations and heat transport occurs in chaotic bursts, whereas for higher $Pr$ the flow does not burst and vertically-sheared thermal plumes dominate the structure of the flow at all times. In both regimes, once the zonal flow sets in, its net effect is the decrease of the vertical heat transport. Interestingly, bursts and sheared convective plumes have also been found in two-dimensional simulations of fingering convection \citep[e.g.,][]{2015ApJ...815...42G,2019JFM...858..228X, 2019ApJ...879...60G}, and as with zonal flows in thermal convection, they decrease the vertical transport of heat and solute.

The effects discussed above have also raised the question of whether two-dimensional simulations are appropriate to model convection in non-rotating systems or in systems where zonal flows are not expected to occur. In particular, the suppression of the vertical transport by zonal flows might be a problem for studies of turbulent mixing at convective boundaries, and layer formation in double-diffusive convection \citep[where in the latter there is good agreement between 2D simulations and laboratory experiments, see, e.g., ][]{molemaker_dijkstra_1997}. However, recent work by \citet{2014PhFl...26e4104F} and \citet{2020JFM...905A..21W} has shown that for the case of pure thermal convection, zonal flows are not sustained in two-dimensional simulations as long as the computational domain has a large enough aspect ratio.


We report two-dimensional simulations of convection driven by a constant heat flux at the top boundary in a stable fluid with a solute gradient. This configuration is particularly useful to study the onset and evolution of zonal flows. The solute gradient stabilizes the fluid against overturning convection in the whole fluid domain, leaving a convective layer whose thickness (aspect ratio) increases (decreases) with time. This allows the study of convection for a wide range of $Ra$ and aspect ratio using a smaller grid of simulations. Further, it provides an opportunity to study the transition to sheared convection once the zonal flow arises if they do. It is worth mentioning that a similar flow can be achieved with just a single scalar determining the density, as in the penetrative convection experiments by \citet{deardorff_willis_lilly_1969}, where the fluid was initially stably stratified with temperature and suddenly heated from below. However, our original interest in this setup was in understanding the speed at which convection propagates into a stable layer, as reported in \citet{2020PhRvF...5l4501F}. Here we use a similar setup to study the onset of zonal flows. 

We perform simulations at $Pr$ ranging from 0.1 to 7 in order to explore the bursting and non-bursting regimes observed in thermal convection. This extends the previous work on the bursting regime of shearing convection to lower $Pr$. We also perform simulations with fluid domains of different aspect ratio to reveal whether zonal flows and their shear effects appear in domains of large aspect ratio. 

The paper is organised as follows. Sect.~\ref{sect_model} presents the model and the numerical code used to perform the simulations. In Sect.~\ref{sect_square}-\ref{sect_transport} we study the onset of zonal flows and the bursting and non-bursting regimes in simulations with aspect ratio of one (i.e., $L=H$). In Sect.~\ref{sect_wider} we show that strong horizontal flows and their effects vanish when the width of the domain is increased ($L\geq 2H$). We conclude in Sect.~\ref{sec_conclusions}.

\section{Model and numerical method} \label{sect_model}

We study the onset and evolution of zonal flows in a two-dimensional convective layer that grows inward by incorporating fluid from below. These simulations are based on our previous study \citep{2020PhRvF...5l4501F} that was focused on the rate at which the convection zone grows inwards. However, in the previous work we deliberately excluded those simulations that developed shearing convection. In the simulations, we model the fluid under the Boussinesq approximation \citep{1960ApJ...131..442S}, which is appropriate when the density variations are small respect to the background density ($\rho/\rho_0\ll 1$).  The fluid domain is a Cartesian box in two-dimensions ($x$,$z$) of width $L$ and height $H$,  with periodic boundary conditions in the horizontal direction. The top and bottom boundaries are impermeable and stress-free, with no composition flux through them, no heat flux at the bottom, and a constant heat flux $F_0$ at the top. Initially, the fluid starts with a uniform temperature and a constant solute gradient $d\overline S_0/dz  = - \delta S_0 / H < 0 $, such that the solute concentration is two times larger at the bottom compared to the top.

The governing equations are
\begin{gather}
\nabla \cdot \bm{v} = 0\, ,\\
\dfrac{\partial T}{\partial t}  = - (\bm{v}\cdot \nabla)\,T + \kappa_T\nabla^2 T\, ,\\
\dfrac{\partial S}{\partial t} = - (\bm{v}\cdot \nabla)\,S + \kappa_S\nabla^2 S\, ,\\
\dfrac{\partial \bm{v}}{\partial t} = - (\bm{v}\cdot \nabla)\,\bm{v} - \dfrac{\nabla P}{\rho_0} + \left(\dfrac{\rho}{\rho_0}\right)\bm{g} + \nu\nabla^2 \bm{v}\,,
\end{gather}
where $T$, $S$ and $\rho = \rho_0(\beta S - \alpha T)$ are the temperature, solute, and density perturbations, respectively (with $\rho_0$ the background density, and $\beta$ and $\alpha$ the coefficients of solute and thermal contraction-expansion, respectively), $\bm{v} = (u,w)$ is the velocity field (being $u$ and $w$ the $x$ and $z$ component, respectively), $P$ is the pressure fluctuation resulting from the fluid motions, $\bm{g}$ is the acceleration due to gravity, and $k = \rho_0 c_P \kappa_T$ is the thermal conductivity (where $\kappa_T$ is the thermal diffusivity, and $c_P$ is the specific heat capacity at constant pressure). The parameter values used in the simulations can be found in Table 1 of \cite{2020PhRvF...5l4501F}.

The boundary conditions are
\begin{gather}
w\,\big\vert_{z=0,H} = 0\,, \hspace{0.2cm} \dfrac{\partial  u}{\partial z}\,\bigg\vert_{z=0,H} = 0\,, \hspace{0.2cm} \dfrac{\partial S}{\partial z}\,\bigg\vert_{z=0,H} = 0\,,\\
\dfrac{\partial T}{\partial z}\,\bigg\vert_{z=0} = 0\, , \hspace{0.2cm} \dfrac{\partial T}{\partial z}\,\bigg\vert_{z=H} = -\dfrac{F_0}{k}\, .
\end{gather}
It is worth clarifying that we choose zero flux boundary conditions for solute to ensure conservation within the box.
The inconsistency between the initial solute gradient (uniform across the fluid) and the zero flux boundary conditions for solute does not have a significant effect on our calculations because the running time of the simulations is at most $0.5\%$ of the time that takes for solute to diffuse across the box. Further, although the initial gradient is eroded near the top and bottom boundaries, it only slightly affects the bottom boundary. The convective motions near the top rapidly mix the initial gradient, making the solute concentration uniform everywhere inside the convection zone ($\partial S/\partial z = 0$, including the top boundary).

Following Sect. 2 in \citep{2020PhRvF...5l4501F}, we non-dimensionalize the variables using as characteristic length and time scales the box height, $H$, and the thermal diffusion time across the box, $t_{\rm diff} = H^2/\kappa_T$. Note that after this choice the velocity scale is $v_{\mathrm{diff}} = \kappa_T/H$. Further, solute concentration is measured in units of the initial solute contrast across the box, $\delta S_0$, and the temperature unit is written in terms of the imposed flux as $F_0 H/k$. The resulting dimensionless equations (with the dimensionless variables written with a tilde on the top) are
\begin{gather}
\nabla \cdot \bm{\tilde v} = 0\, ,\\
\dfrac{\partial \tilde T}{\partial \tilde t}  = - (\bm{\tilde v}\cdot \nabla)\,\tilde T + \nabla^2 \tilde T\, ,\\
\dfrac{\partial \tilde S}{\partial \tilde t} = - (\bm{\tilde v}\cdot \nabla)\,\tilde S + \tau\nabla^2 \tilde S\, ,\\
\dfrac{\partial \bm{\tilde v}}{\partial \tilde t} = - (\bm{\tilde v}\cdot \nabla)\,\bm{\tilde v} - \nabla \tilde P + \mathcal{R}_T Pr \left[ \tilde{T} -
\left(\frac{F_0}{F_\mathrm{crit}}\right)^{-1} \tilde{S} \right]\bm{\hat z} + Pr\nabla^2 \bm{\tilde v}\,,
\end{gather}
where $F_{\rm crit} \equiv k (\beta/\alpha)\, |d\overline S_0/dz|$ is the diffusive heat flux (through the entire box) for which the fluid is marginally stable against convection.
The boundary conditions in dimensionless form read
\begin{gather}
\tilde w\,\big\vert_{\tilde z=0,1} = 0\,, \hspace{0.2cm} \dfrac{\partial  \tilde u}{\partial \tilde z}\,\bigg\vert_{\tilde z=0,1} = 0\,, \hspace{0.2cm} \dfrac{\partial \tilde S}{\partial \tilde z}\,\bigg\vert_{\tilde z=0,1} = 0\,,\\
\dfrac{\partial \tilde T}{\partial \tilde z}\,\bigg\vert_{\tilde z=0} = 0\, , \hspace{0.2cm} \dfrac{\partial \tilde T}{\partial \tilde z}\,\bigg\vert_{\tilde z=1} = -1\, .
\end{gather}

\begin{table}
\centering
\caption{Dimensionless parameters used in the simulations. The first and second columns correspond to the diffusivity ratio and Prandtl number, respectively. The third column contains the input cooling flux at the top boundary, and the last two columns contain the modified Rayleigh numbers for temperature and solute, respectively.} \label{Table_a1}
\begin{tabular}{@{}llllll@{}}
\hline
$\#$ & $\tau$ & $Pr$ & $F_0/F_{\rm crit}$ & $\mathcal{R}_T$ & $\mathcal{R}_S$ \\
\hline
1 & 0.1 & 0.1 & 5.4 & $4\times 10^{12}$ & $7.5\times 10^{11}$ \\ 
2 & 0.1 & 0.1 & 10.8 & $8\times 10^{12}$ &  $7.5\times 10^{11}$ \\ 

3 & 0.1 & 0.5 & 5.4 & $8\times 10^{11}$ & $1.5\times 10^{11}$\\ 
4 & 0.1 & 0.5 & 10.8 & $1.6\times 10^{12}$ &  $1.5\times 10^{11}$\\ 

5 & 0.1 & 1 & 5.4 & $4\times 10^{11}$ &  $7.5\times 10^{10}$ \\ 
6 & 0.1 & 1 & 10.8 & $8\times 10^{11}$ &  $7.5\times 10^{10}$ \\ 
7 & 0.1 & 7 & 5.4 & $5.76\times 10^{10}$ & $1.06 \times 10^{10}$ \\ 
8 & 0.1 & 7 & 10.8 & $1.15\times 10^{11}$ & $1.06 \times 10^{10}$ \\ 
\hline
\end{tabular}
\end{table}

The dimensionless parameters that control the simulations are $F_0/F_{\rm crit}$, the Prandtl number ($Pr$), the diffusivity ratio ($\tau$) and a modified Rayleigh number ($\mathcal{R}_T$), defined respectively as
\begin{align}
&\dfrac{F_0}{F_{\rm crit}} = F_0\left(k\dfrac{\beta}{\alpha}\dfrac{\delta S_0}{H}\right)^{-1}\, ,\\
&Pr = \dfrac{\nu}{\kappa_T}\, ,\\
&\tau = \dfrac{\kappa_S}{\kappa_T}\, ,\\
&\mathcal{R}_T = \dfrac{\alpha g H^3}{\kappa_T \nu }\left(\dfrac{F_0 H}{k}\right),
\end{align}
where $\nu$, and $\kappa_S$ are the viscous and solute diffusivity, respectively, and $g$ is the magnitude of the acceleration due to gravity.  As we note in \citep{2020PhRvF...5l4501F}, the product $\mathcal{R}_T (F_0/F_{\rm crit})^{-1}$ can be re-written as
\begin{align}
\mathcal{R}_T \left(\dfrac{F_0}{F_{\rm crit}}\right)^{-1} = \mathcal{R}_S = \dfrac{\beta g H^3 \delta S_0}{\kappa_T \nu },
\end{align}
which resembles the traditional Rayleigh-number of thermal convection when $\delta S_0 = (\alpha/\beta)\, \delta T_0$. Table \ref{Table_a1} provides a list with the numerical values of the dimensionless parameters used in our simulations.

The governing differential equations are solved with the Dedalus spectral code \citep{2020PhRvR...2b3068B} on a Chebyshev (vertical) and Fourier (horizontally-periodic) domain in which the physical grid dimensions are 3/2 the number of modes. Based on a resolution study (Appendix~\ref{sect_res_study}), we use 1024 modes in each direction.  
The system is initialized by adding random noise of small amplitude to the temperature perturbation at the top boundary. The interaction of the initial noise and the cooling flux $F_0$ at the top boundary quickly forms a convective layer that grows inwards.

\section{Analysis and results}

\subsection{Development of zonal flow driven by convection} \label{sect_square}

\begin{figure*}
\hspace{0.7cm}\includegraphics[width=6cm]{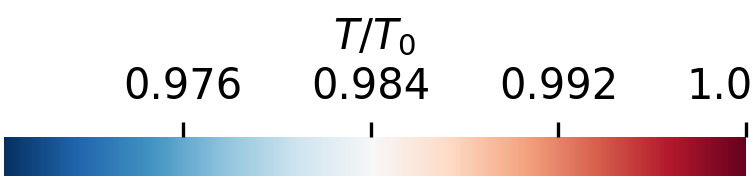}\\
\includegraphics[width=\textwidth]{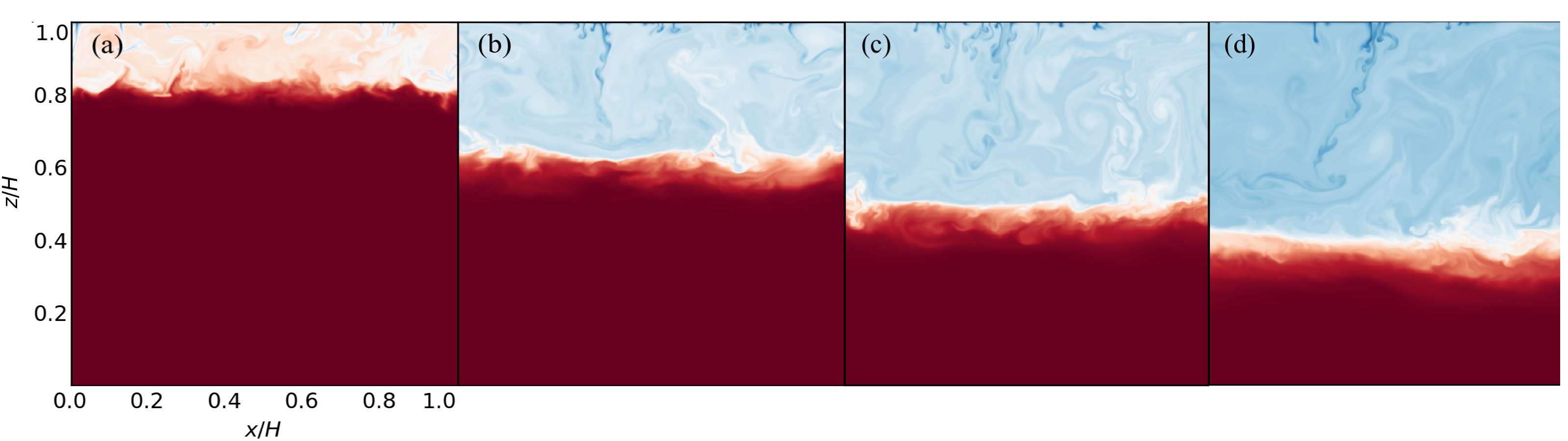}
\caption{Instantaneous snapshots of the temperature field for the run using $Pr=0.5$ and $F_0/F_{\rm crit} = 10.8$. Panels (a)-(c) show snapshots during the early evolution of the convective layer, when the flow is dominated by cellular motions. Panel (d) shows a snapshot when convective plumes become tilted due to advection by horizontal flows. For better visualization we show labels and ticks just in panel (a). All panels share the same color scale. } \label{fig_growth}
\end{figure*}

As soon as the heat flux at the top boundary turns on, a thermal boundary layer develops and becomes convective. The recently-formed convective layer is composed of an array of convective plumes that exhibit cellular motions of horizontal size approximately the height of the layer.  As the layer increases its thickness, both the aspect ratio of the flow and the number of convective cells within the layer decrease (Fig.~\ref{fig_growth} a-c). Once the growing convective layer reaches a critical size, convective plumes become tilted respect to the vertical (Fig.~\ref{fig_growth}d).

\begin{figure*}
\centering
\includegraphics[width=8cm]{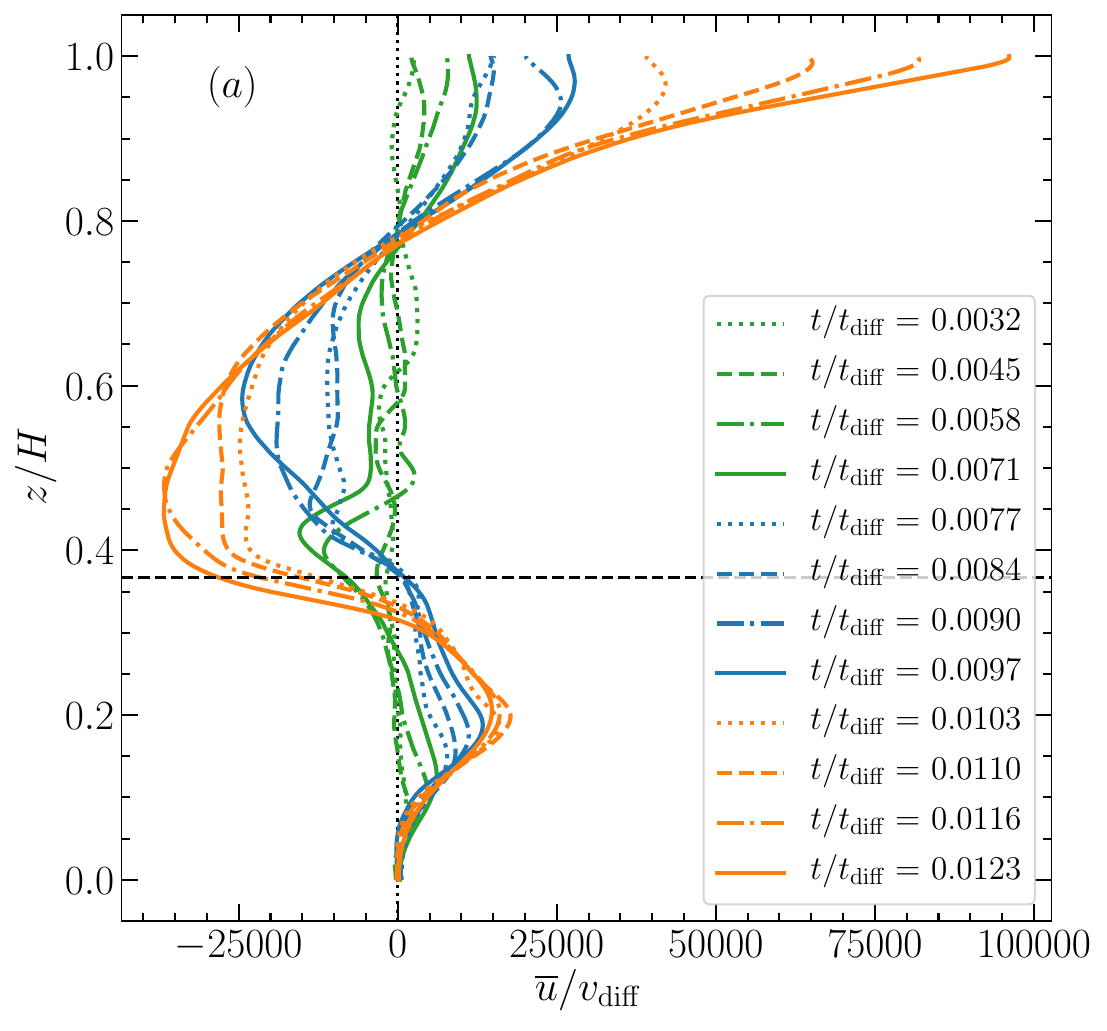} \hspace{0.01cm}\includegraphics[width=7.7cm]{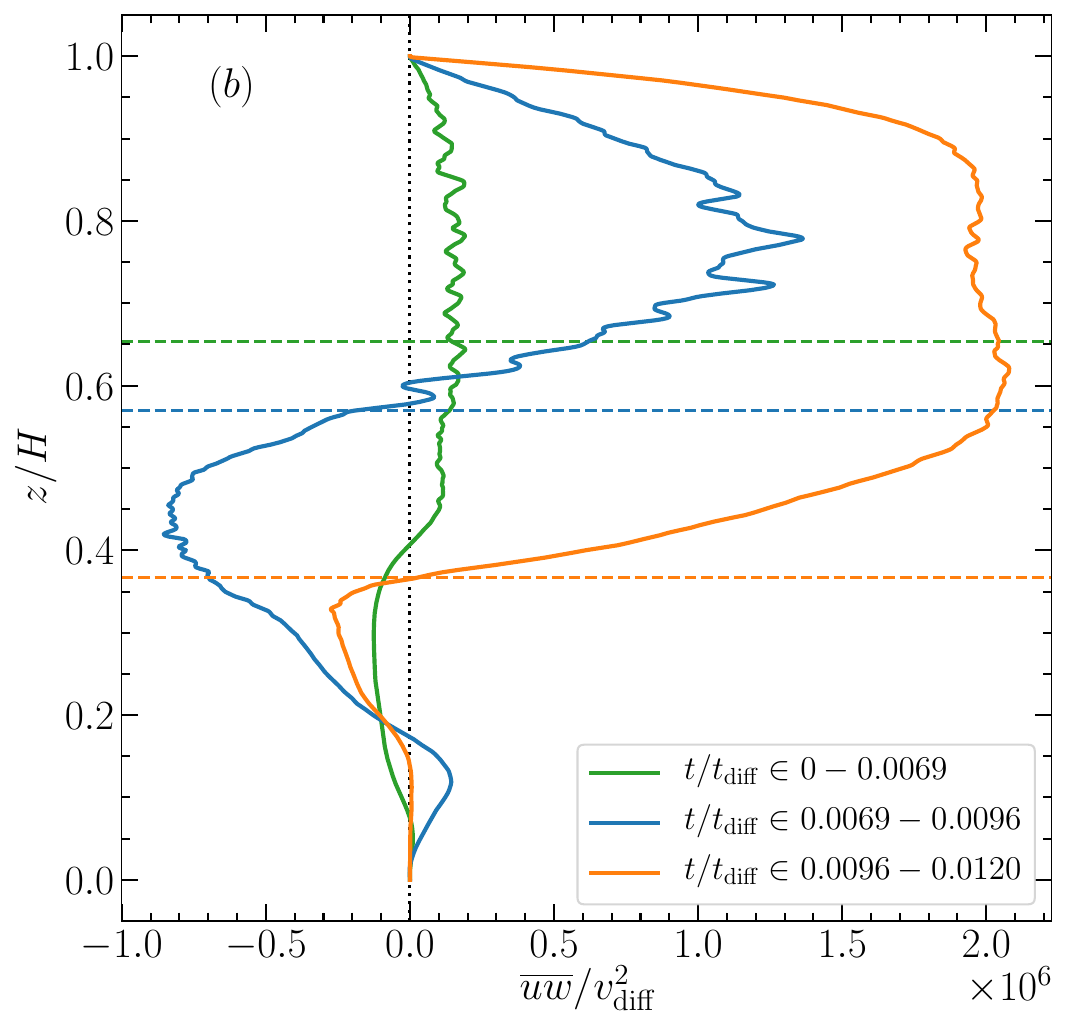}
\caption{Panel (a): Snapshots of the horizontal velocity profiles normalized to $v_{\rm diff}$. Different colors and lines were used to illustrate the transition from zero mean flow to a large scale horizontal flow vertically sheared. The dashed horizontal line denotes the location of the convective boundary at $t/t_{\rm diff} = 0.0123$. Panel (b): Snapshots of Reynolds stress profiles normalized to $v^2_{\rm diff}$. Since the Reynolds stress exhibits large fluctuations over time, we time-average the profiles over the intervals shown in the legends. The horizontal dashed lines denote the location of the convective boundary at the upper edge of the time intervals in the legends. In both panels, results are shown for the run using $Pr=0.5$ and $F_0/F_{\rm crit} = 10.8$.} \label{fig_development_shear}
\end{figure*}

We compute horizontally-averaged profiles of the horizontal velocity and the Reynolds stresses associated with the tilted plumes ($\overline{u}$ and $\overline{uw}$, respectively, where $\bar{\cdot}$ denotes the average over the horizontal direction) at different times during the evolution of the flow. Fig.~\ref{fig_development_shear} shows results for the run using $Pr = 0.5$ and $F_0/F_{\rm crit} = 10.8$. We find that the mean horizontal velocity evolves from being roughly zero to a strong flow that is vertically sheared, directed to the right at the top of the convection zone and to the left at the bottom (Fig.~\ref{fig_development_shear} a). Note that the magnitude of the zonal flow near the bottom of the convective layer is much weaker than its value at the top. A possible cause for this is the interaction with the initially motionless fluid below the convection zone which slows down the mean flow near the convective boundary. On the contrary, the top of the convection zone has no stresses and therefore the mean flow there does not slow down. The profiles of the Reynolds stress are consistent with the enhancement of the zonal flow (Fig.~\ref{fig_development_shear} b). The stresses exhibit random behaviour during the early evolution of the flow, with magnitudes oscillating between negative and positive values. As the zonal flow develops, the Reynolds stress becomes positive inside the convection zone, with a negative (positive) vertical gradient in the upper (lower) half of the layer.  This indicates that positive $x$-momentum is being transported upward.

It is worth mentioning that we do not find any trend regarding the direction of the zonal flow; in some of the simulations it goes to $+x$ ($-x$) at the top (bottom) boundaries of the convection zone, and vice-versa. However, we find for all the cases that once the zonal flow sets in, its direction is never reversed during its evolution.

\subsection{Energy evolution of the zonal flow}

In the following we analyze the temporal evolution of the kinetic energy of the flow. We define the volume-averaged horizontal, vertical, and total kinetic energy of the flow as $E_x = \frac{1}{2}\rho_0 \langle u^2 \rangle$, $E_z = \frac{1}{2}\rho_0 \langle w^2 \rangle$, and $E = E_x + E_z$, respectively, where $\langle \cdot \rangle$ denotes the average over the entire domain. Using mixing length theory, we also estimate the kinetic energy associated with convective motions as 
\begin{equation}
E_{\rm conv} \sim \dfrac{1}{2} \rho_0 v_{\rm conv}^2 \sim \dfrac{1}{2} \rho_0  g \alpha \delta T \ell_{\rm mix}\text{,} \label{eq_E_conv}
\end{equation}
where $\ell_{\rm mix}$ is the mixing length, and $\delta T$ is the temperature fluctuation that drives convection. We write the energy flux carried by convective motions as
\begin{equation}\label{eq:mixlength2}
F_{\rm conv} \sim \rho_0 c_P v_{\rm conv} \delta T\, 
\end{equation}
from which we estimate the temperature fluctuation as
\begin{equation}
\delta T \sim \left(\dfrac{F_{\rm conv}}{\rho_0 c_P}\right)^{2/3} \left(\dfrac{1}{g\alpha \ell_{\rm mix}}\right)^{1/3}\, .  \label{eq_dT}
\end{equation}
Combining Eqs. \eqref{eq_E_conv} and \eqref{eq_dT} the kinetic energy associated with convective motions is determined by $F_{\rm conv}$ and $\ell_{\rm mix}$
\begin{equation}
E_{\rm conv} \sim \dfrac{1}{2} \rho_0 \left(\dfrac{\alpha g F_{\rm conv} \ell_{\rm mix}}{\rho_0 c_P}\right)^{2/3}\, . \label{eq_conv_e}
\end{equation}
We measure the flux carried by convection from the simulations (see Sect. \ref{sect_transport}) and use its volume-averaged value as the representative magnitude of $F_{\rm conv}$. To estimate the mixing length, before the onset of the zonal flow we set $\ell_{\rm mix} = h$, where $h$ is the size of the convection zone, measured as the distance between the top boundary and the location where the solute concentration varies at most by $5\%$ respect to its value at the top boundary. As we discuss later, once the zonal flow arises, the fluid elements that sink from the top boundary get advected by the flow, reducing the effective mixing length. As a rough estimation, we assume that a fluid element gets dispersed when the change in its horizontal velocity is the same as its vertical velocity. This condition gives $
\ell_{\rm mix} \sim |w/(\partial u/\partial z)|$. To interpolate between the two regimes and for numerical estimations, at each time we compute the mixing-length as $\ell_\mathrm{mix} = \mathrm{min}(h,\langle |w/(\partial u/\partial z)|\rangle)$.

\begin{figure}
\centering
\includegraphics[width=8cm]{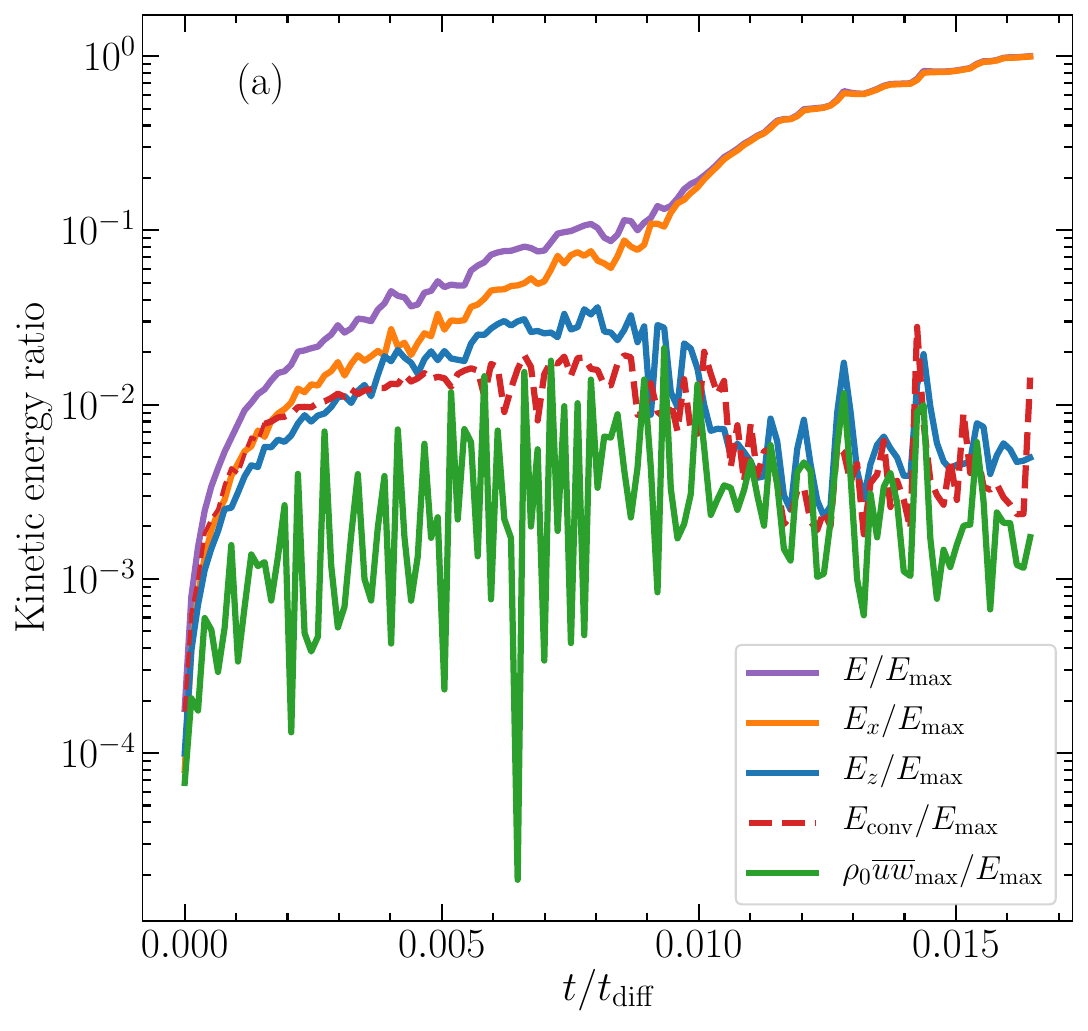}\hspace{0.01cm}
\includegraphics[width=8cm]{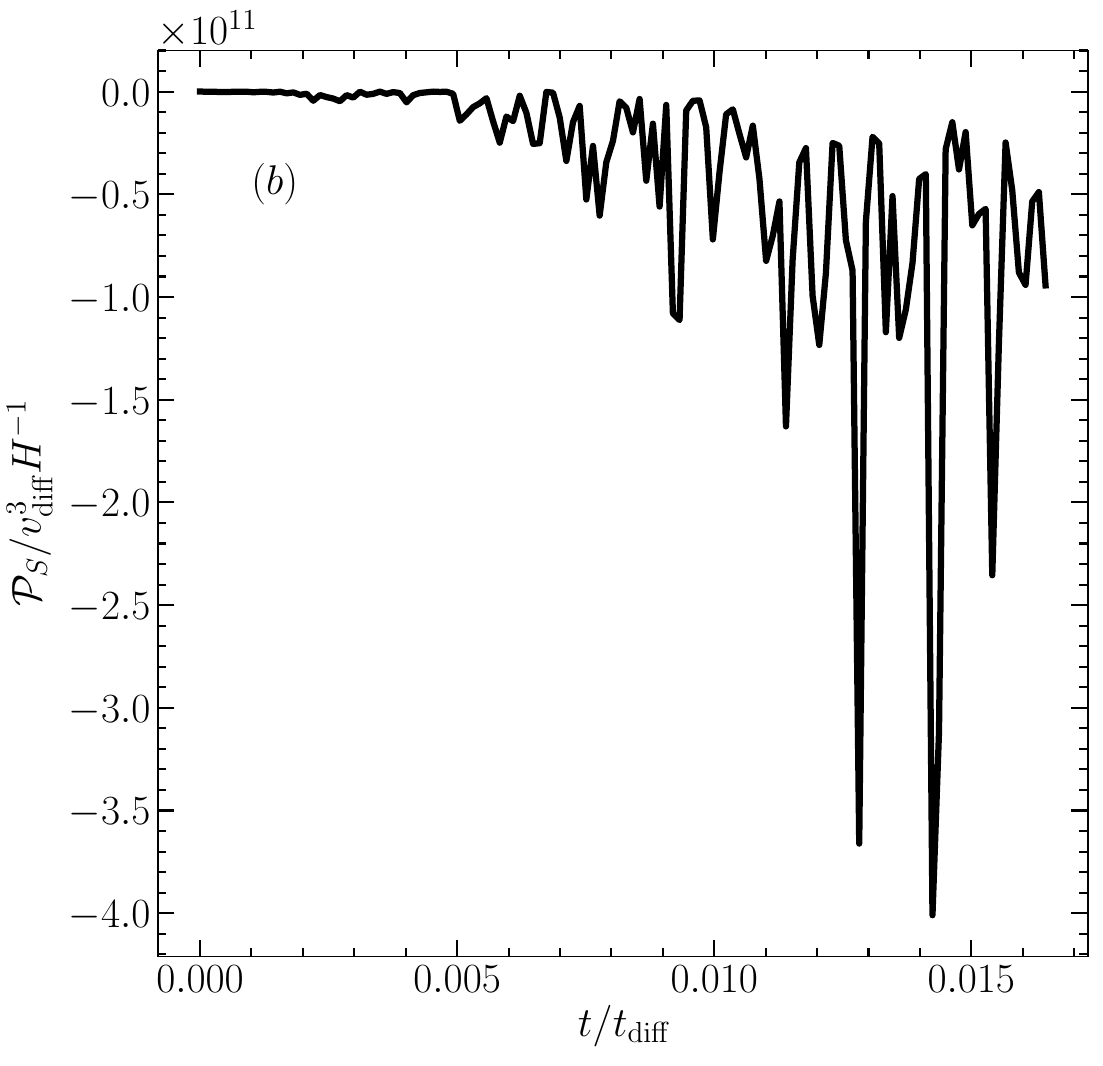}
\caption{Panel (a): Time series of the volume-averaged kinetic energies and maximum Reynolds stress (see text for their definitions). All the curves are normalized to the maximum value of the total kinetic energy, i.e., $E_{\rm max} = E(t/t_{\rm diff} \approx 0.017)$. Panel (b): Time series of the shear production $\mathcal{P}_S$ (Eq. \ref{eq_shear_p}) divided by $v^3_{\rm diff} H^{-1}$ (the scales for velocity and length). Note that $\mathcal{P}_S <0$ indicates a net transfer of energy from the convective motions to the mean zonal flow. In both panels the results are shown for the run using $Pr=0.5$ and $F_0/F_{\rm crit} = 10.8$.} \label{fig_kinetic_energy}
\end{figure}

Figure \ref{fig_kinetic_energy}(a) shows time series of the kinetic energies described above, and of the maximum Reynolds stress in the convection zone. As in Figs. \ref{fig_growth} and \ref{fig_development_shear}, the results are shown for the run using $Pr=0.5$ and $F_0/F_{\rm crit} = 10.8$. We distinguish three phases. First, the horizontal and vertical kinetic energies track each other until $t/t_{\rm diff} \approx 0.007$. This is expected when the flow pattern is dominated by an array of plumes going down and up, whose horizontal scale is of the same order as the layer depth. Second, from $t/t_{\rm diff} \approx 0.007$ to $t/t_{\rm diff} \approx 0.01$, the horizontal kinetic energy continues to grow but not the vertical. At this point, the flow pattern contains plumes that are tilted from the vertical, and the horizontal flow starts to dominate the total energy. Third, for $t/t_{\rm diff} > 0.01$, the vertical kinetic energy starts to decrease and the horizontal one has a large increase, dominating the total kinetic energy. Note that our analytic estimation of the convective kinetic energy (Eq. \ref{eq_conv_e}) is consistent with $E_z$ (as expected). We observe that during the first two phases the maximum value of the Reynolds stress increases with time, reinforcing the horizontal fluids motions by transporting momentum upwards. Further, we observe that the vertical kinetic energy sets the maximum possible value of the Reynolds stress. This suggests that energy is transferred from the vertical convective motions to the zonal flow. The peaks observed for $t/t_{\rm diff} > 0.01$ are the result of quasi-periodic convective bursts and we discuss them in Sect.~\ref{sect_transport}.

The transfer of energy from the convective motions to the mean zonal flow can be determined by decomposing the total energy as
\begin{equation}
E = E_{\rm zonal} + E' = \dfrac{1}{2}\rho_0\left(\langle u^2_{\rm zonal} \rangle + \langle \bm{v}^{\prime 2} \rangle \right)\, ,
\end{equation}
where we decompose the total velocity into the mean zonal flow and the fluctuations from it
\begin{equation}
\bm{v}(x,z,t) = \overline{u}_{\rm zonal}(z,t) \bm{\hat{x}} + \bm{v}^{\prime} \,
\end{equation}
Note that $\bm{v}^{\prime} = (u^{\prime},w^{\prime})$,  $\bm{v} = (u,w)$, therefore $u = \overline{u}_{\rm zonal} + u'$ and $w = w'$. Note that under this Reynolds decomposition, $\overline{u} = \overline{u}_{\rm zonal}$ and $\overline{uw} = \overline{u^{\prime}w^{\prime}}$.  The evolution of the zonal kinetic energy is given by
\begin{equation}
{d E_\mathrm{zonal}\over dt} = - \rho_0\mathcal{P}_S - \rho_0\mathcal{D}\,
\end{equation}
where $\mathcal{P}_S$ and $\mathcal{D}$ are the shear production and viscous dissipation rate respectively, defined as  
\begin{equation}
\mathcal{P}_S = \bigg\langle-u'w'\dfrac{\partial\overline{u}}{\partial z} \bigg \rangle \, , \hspace{0.5cm} \mathcal{D} = \nu\,\bigg\langle \bigg|\dfrac{\partial\overline{u}}{\partial z}  \bigg|^2 \bigg \rangle \, . \label{eq_shear_p}
\end{equation}
Equation \eqref{eq_shear_p} shows that shear production from fluctuations increases the zonal kinetic energy, whereas viscosity dissipates the zonal flow. We can infer from Fig.~\ref{fig_development_shear}
that the shear production is negative inside the convection zone. As a double check, we show in Fig.~\ref{fig_kinetic_energy}b time series of $\mathcal{P}_S$ and find that it is negative as the zonal flow develops over time, indicating a net transfer of energy from the convective motions to the zonal flow.

\begin{figure}
\centering
\includegraphics[width=8cm]{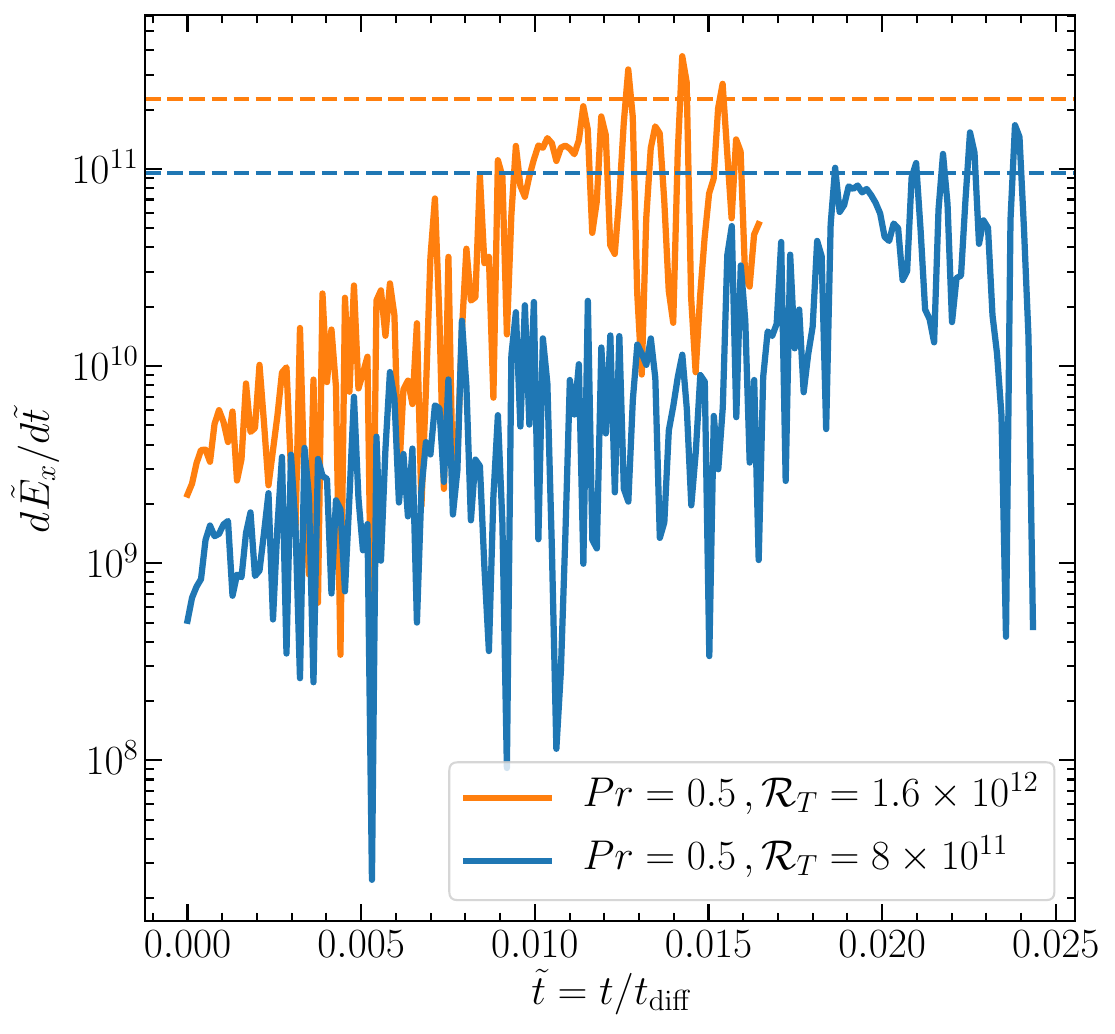}
\caption{Time series of the rate of change of horizontal kinetic energy (dimensionless), $d\tilde{E}_x/d\tilde{t}$, for the runs using $Pr = 0.5$ (fixed), and $F_0/F_{\rm crit} = 5.4$ ($\mathcal{R}_T = 8\times 10^{11}$ in blue) and $F_0/F_{\rm crit} = 10.8$ ($\mathcal{R}_T = 1.6\times 10^{12}$ orange). We measure $d\tilde{E}_x/d\tilde{t}$ directly from the simulations using the values of $\langle \tilde{u}^2\rangle$ as a function of time. The dashed-horizontal lines correspond to the order of magnitude estimation in Eq. \eqref{eq_exdot_dimensionless}.}  \label{fig_odm}
\end{figure}


We can use mixing length theory to estimate the expected size of $\mathcal{P}_S$. From equations (\ref{eq_E_conv}) and (\ref{eq:mixlength2}), we expect convective velocities $w^\prime\approx v_\mathrm{conv}$ where $v_\mathrm{conv}^3\approx \ell_\mathrm{mix}g\alpha F_\mathrm{conv}/\rho_0 c_P$, with a mixing length in the zonal flow of $\ell_\mathrm{mix}\approx w/(\partial u/\partial z)$. The heat flux in the convection zone is $F_{\rm conv}\approx (1/2)F_0$, where the factor of $1/2$ comes from the fact that the fluid is cooling down at a constant rate everywhere inside the convection zone (as we discuss in Sect. \ref{sect_transport}).
Therefore, we estimate
\begin{equation}
-\rho_0\mathcal{P}_S \sim \eta \rho_0w^{\prime 2} {d\bar{u}\over dz}\sim \eta \rho_0{w^{\prime 3}\over \ell_\mathrm{mix}} \sim {1\over 2} \eta \left(\dfrac{g\alpha F_0}{c_P}\right)\, , \label{eq_exdot}
\end{equation}
where $\eta = \langle u^{\prime} w^{\prime}\rangle/\langle w^{\prime 2}\rangle$ measures the relative size of the horizontal and vertical velocity fluctuations. The fact that the maximum value of Reynolds stress is comparable to the vertical kinetic energy in Fig.~\ref{fig_kinetic_energy} suggests that $\eta$ is of order unity. 
In dimensionless form, Eq. \eqref{eq_exdot} reads
\begin{equation}
\dfrac{d\tilde{E}_\mathrm{zonal}}{d\tilde{t}}\approx -\tilde{\mathcal{P}}_S\sim {1\over 2} \eta Pr \mathcal{R}_T \, , \label{eq_exdot_dimensionless}
\end{equation}
where we have ignored dissipation effects. Computing the temporal average of $\eta$ over the time span defined between the onset of the zonal flow and the end of the simulations, we find $\eta\approx 0.4$ for the runs with $Pr=0.5$ and the two fluxes $F_0/F_{\rm crit} = 5.4$ ($\mathcal{R}_T = 8\times 10^{11}$) and $F_0/F_{\rm crit} = 10.8$ ($\mathcal{R}_T = 1.6\times 10^{12}$). Using this value of $\eta$ together with $Pr$ and $\mathcal{R}_T$ in Eq. \eqref{eq_exdot_dimensionless}, we find it agrees within about a factor of two, see Fig.~\ref{fig_odm}. It would be interesting to further investigate the dependence of $\eta$ on $Pr$ and $\mathcal{R}_T$ for a wider range of these parameters.


\subsection{Aspect ratio and Rayleigh number of the flow}

\begin{figure*}
\includegraphics[width=8cm]{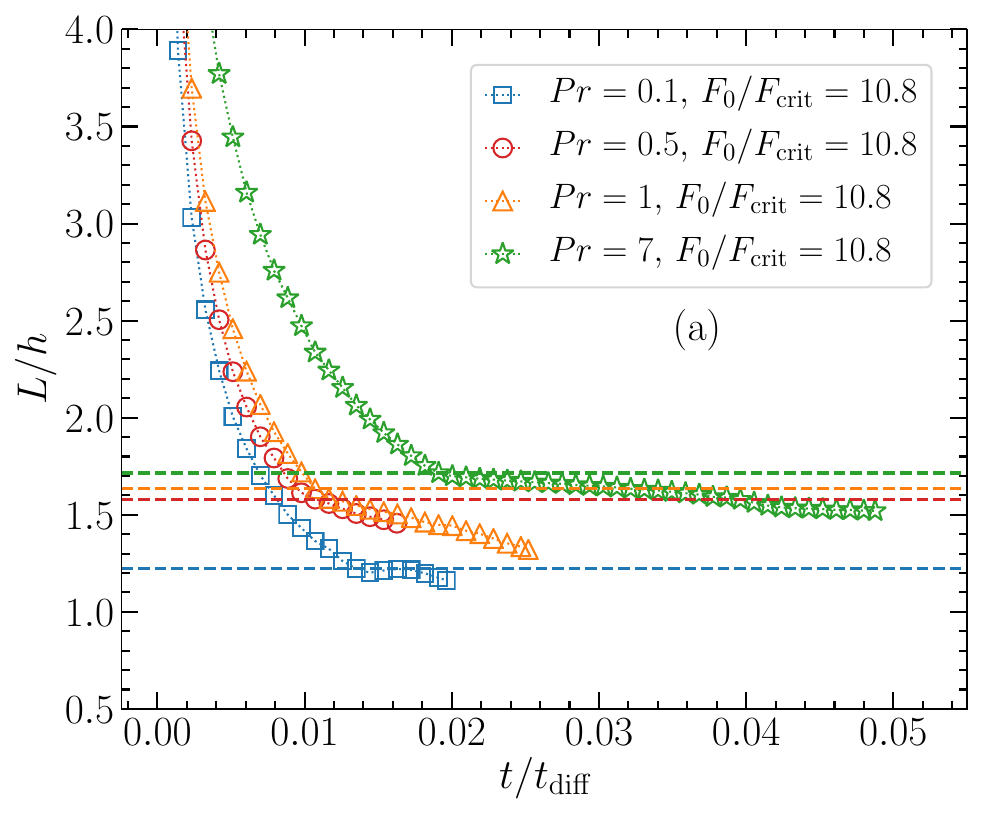}\hspace{0.01cm} \includegraphics[width=8.1cm]{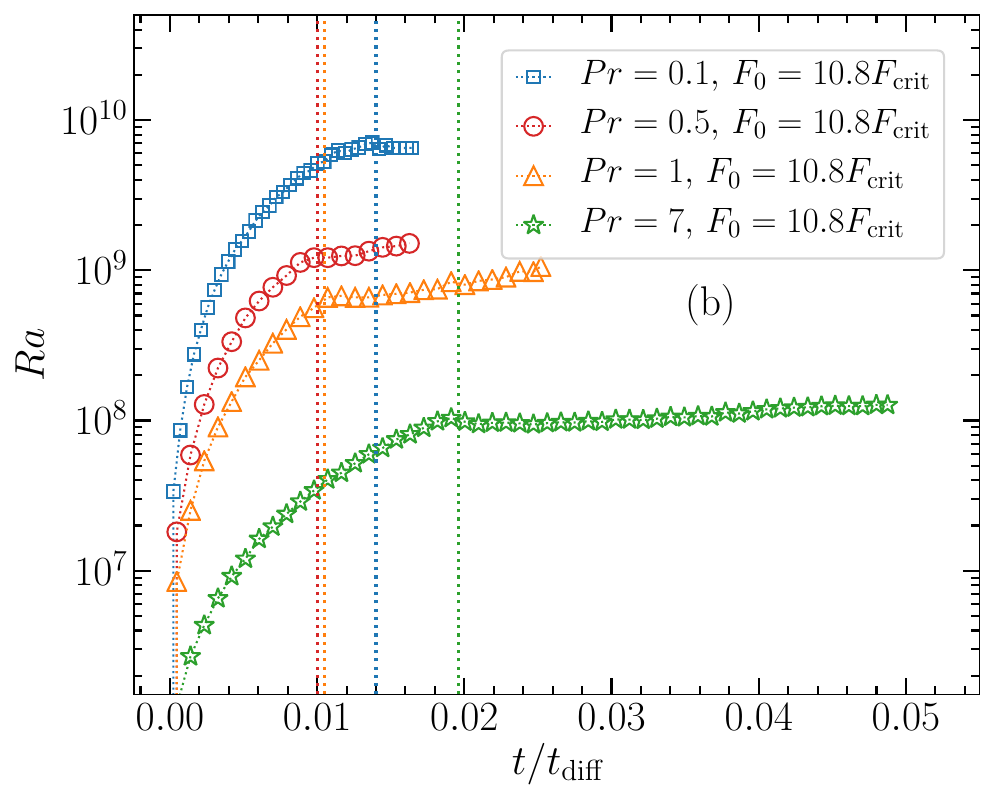}
\caption{Panel (a): Ratio between the width of the domain, $L$, and the thickness of the convective layer, $h$ (i.e., the aspect ratio of the flow within the convective layer) as a function of time. The dashed horizontal lines denote the value of the aspect ratio when the zonal flow sets in. 
Panel (b): Rayleigh number as a function of time. Once the zonal flow sets in (vertical lines), the magnitude of $Ra$ varies slowly with time (flat region). In both panels, the results are shown for simulations using $F_0/F_{\mathrm{ crit}} = 10.8$. Colors and symbols distinguish between simulations at different $Pr$, as shown in the legends. The rest of the simulations behave in a similar way but on different time scales and magnitudes.}\label{fig_aspect_ratio_Ra}
\end{figure*}

\citet{2014PhFl...26e4104F} found that the development of the shear mode in two-dimensional convection depends on the aspect ratio of the convection zone and the magnitude of the Rayleigh-number. In our experiments, we find that zonal flows develop when the aspect ratio of the convection zone lies between $1.2-1.7$, being smaller at $Pr=0.1$ (Fig. \ref{fig_aspect_ratio_Ra}a).

In this problem, convection is driven by the temperature difference across the thermal boundary layer due to the imposed heat flux at the top, and the convective layer grows in time.  This means that the Rayleigh number $Ra$ also increases with time and its magnitude depends mostly on the thickness of the convective layer.  We measure the Rayleigh number of the flow as
\begin{equation}
Ra = \dfrac{\alpha g\, h(t)^3 \Delta T(t)}{\kappa_T \nu}\,    
\end{equation}
where $h(t)$ is the size of the convection zone, and $\Delta T(t) = T_{\rm{CZ}}(t) - T(H,t)$, where $T(H,t)$ is the temperature at the top boundary, and $T_{\rm{CZ}}(t)$ is the temperature of the fluid in the convection zone. We find that once the zonal flow sets in, the convective layer stops growing and $Ra$ saturates at a roughly constant value, $Ra \sim 10^{8},\, 5\times 10^{8},\, 10^9,$ and $5\times 10^{9}$ for $Pr=7,\, 1,\, 0.5$, and 0.1, respectively. To illustrate this behaviour, Fig. \ref{fig_aspect_ratio_Ra}(b) shows the temporal evolution of $Ra$ for the cases using $F_0/F_{\rm crit}=10.8$.

\subsection{Zonal flow and its effect on the vertical transport} \label{sect_transport}

\begin{figure}
\centering
\includegraphics[width=8cm]{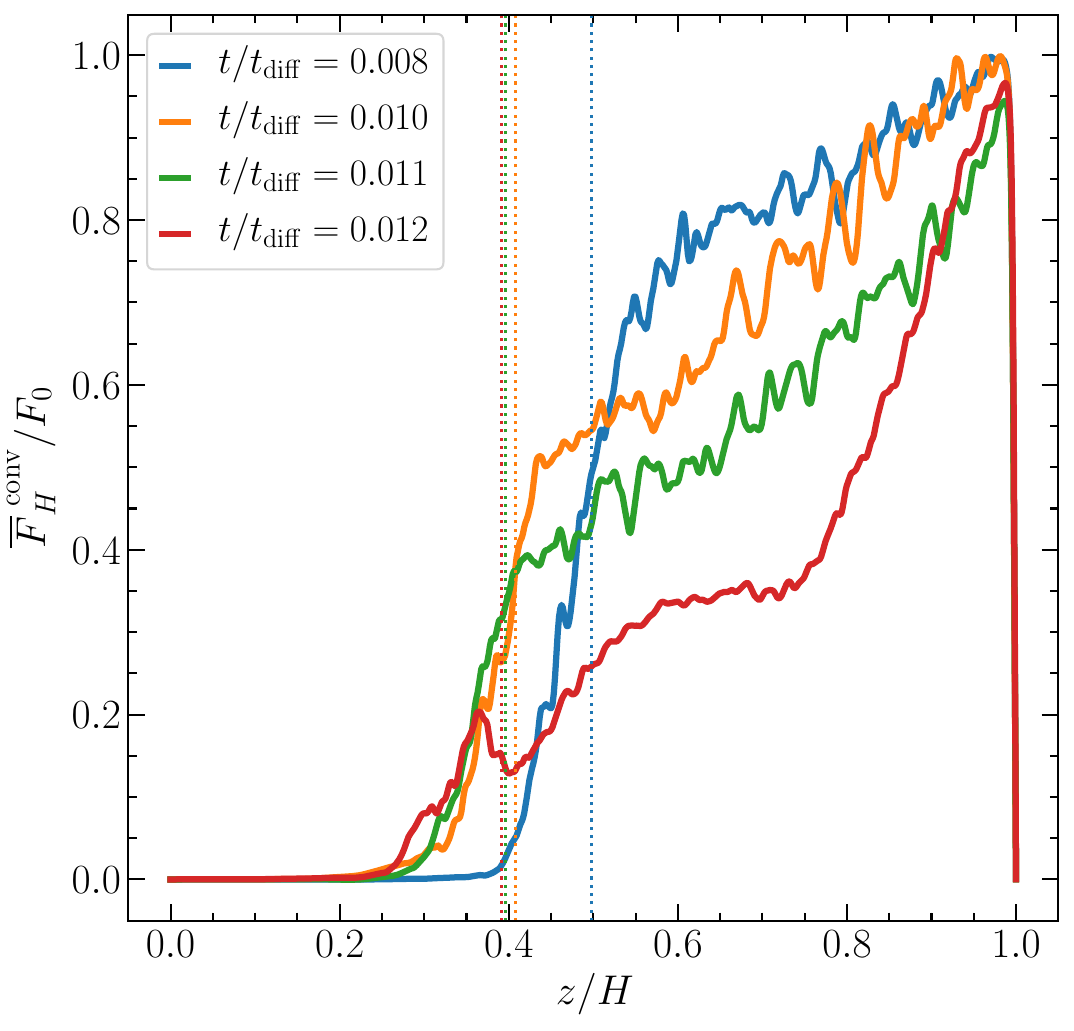}
\caption{Horizontally-averaged profiles of the convective flux divided by the imposed heat flux, $\overline F^{\, \, \rm conv}_H/F_0$. Results are shown at different times  for the run using $Pr=1$ and $F_0/F_{\rm crit} = 10.8$. The vertical lines set the position of the convective boundary at the times when profiles are shown. Colors distinguish between different times. We recall that for this case, from $t/t_{\rm diff}\geq 0.01$ the fluid is affected by the zonal flow.} \label{fig_flux_stuck}
\end{figure}

The stalling of the convective layer growth is explained by the reduced vertical transport of sheared convective plumes. In the absence of strong horizontal flows, kinetic energy in the convective plumes is used to lift and mix fluid from below, increasing the convection zone thickness \citep{molemaker_dijkstra_1997,2020PhRvF...5l4501F}. However, in the situation considered here the zonal flow disperses the convective plumes and takes energy from them. As a result, the mixing length decreases and the convection zone does not grow anymore. The latter is consistent with the horizontally-averaged profiles of the thermal energy flux considering just the contribution from convective motions, $\overline F^{\, \, \rm conv}_H = \rho_0 c_P\overline{wT}$. We observe in Fig. \ref{fig_flux_stuck} that after the zonal flow arises ($t/t_{\rm diff} \geq 0.01$), the convective heat flux near the interface decreases. Further, the position of the interface only slightly changes with time, meaning that the rate at which the outer convection zone grows becomes smaller. We note how the zonal flow affects the shape of the convective flux profiles at $t/t_{\rm diff}=0.012$, changing from being roughly linear within the outer convection zone (the expected profile not affected by the zonal flow, corresponding to a uniform cooling rate) to a profile with roughly two different cooling rates (large in the top half of the layer, $z/H \gtrsim 0.8$, and smaller in the bottom half, $z/H \lesssim 0.8$).

\begin{figure*}
\includegraphics[width=8cm]{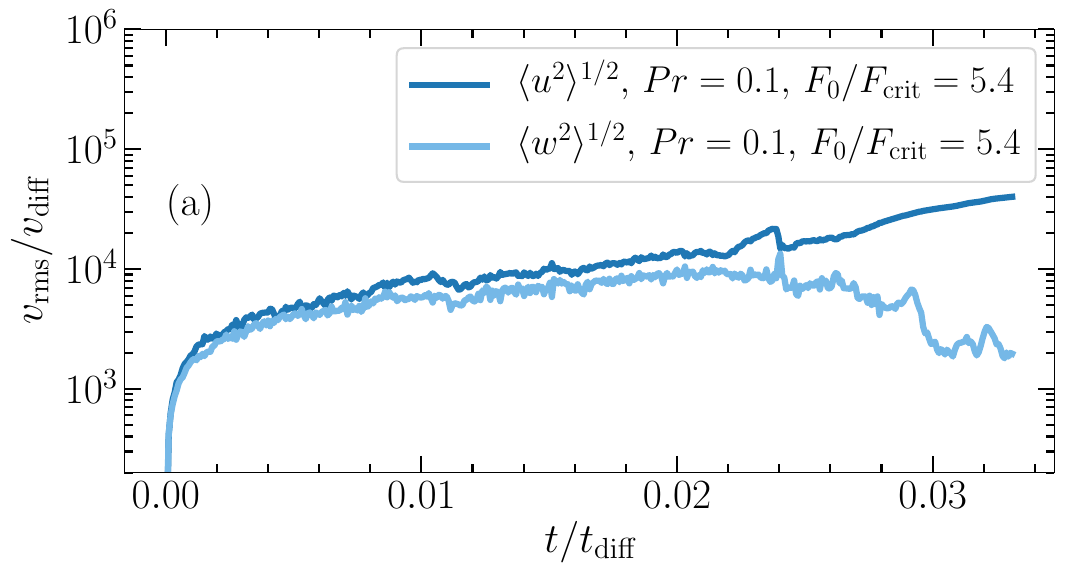}\hspace{0.01cm}
\includegraphics[width=8cm]{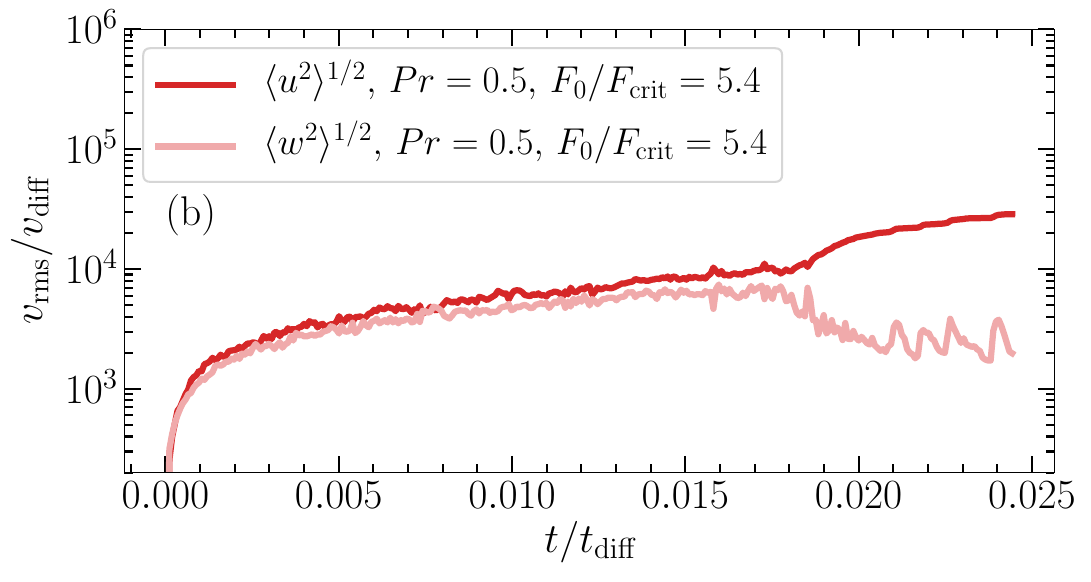}\\
\includegraphics[width=8cm]{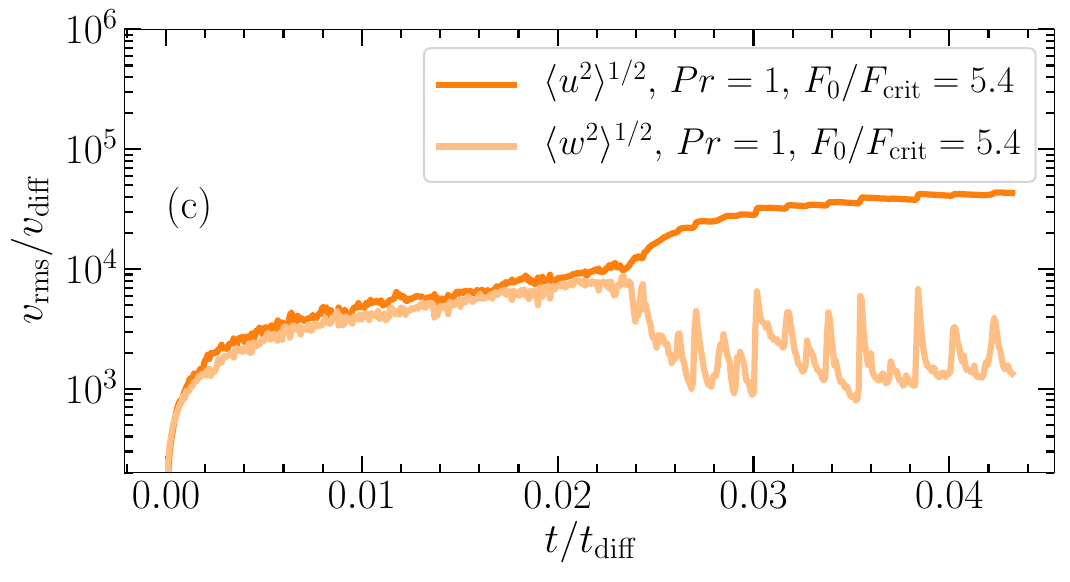}\hspace{0.01cm}
\includegraphics[width=8cm]{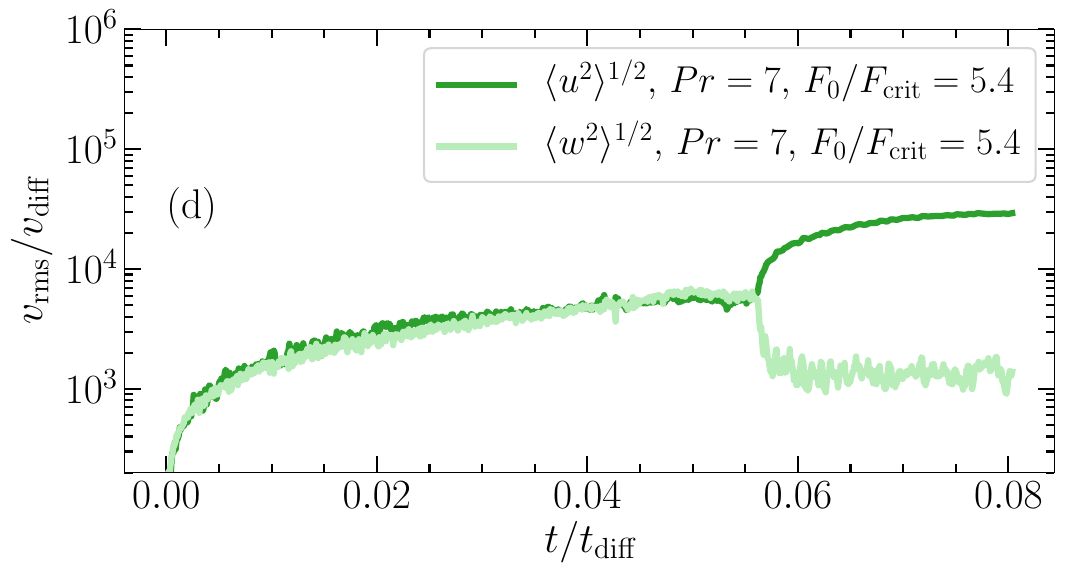}
\caption{Panels (a)-(d): Temporal evolution of the horizontal and vertical rms velocity (dark and light colors) for different $Pr$ using $F_0/F_{\rm crit}=5.4$. Note that panels do not share the same scale in the time axis. The time series for the cases using $F_0/F_{\rm crit}=10.8$ are similar but the instability appears earlier and the magnitude of $\langle u^{2} \rangle^{1/2}$ is higher. }\label{fig_v_f_time}
\end{figure*}

Figure \ref{fig_v_f_time} shows the temporal evolution of $\langle u^{2} \rangle^{1/2}$ and $\langle w^{2} \rangle^{1/2}$ for different $Pr$ using $F_0/F_{\rm crit}=5.4$. We observe that during the early evolution, both rms velocities increase with time having roughly the same magnitude. This is expected since in this stage the convection zone grows and the fluid flow is dominated by cellular motions. However, once the tilting instability begins to operate, $\langle u^{2} \rangle^{1/2}$ increases significantly and $\langle w^{2} \rangle^{1/2}$ suffers a substantial decrease. As we show previously. this behaviour means that a significant fraction of the work done by buoyancy forces is transformed to kinetic energy but enhancing mainly the horizontal fluid motions. Note that for $Pr\leq 1 $, the time series of the rms velocities exhibit oscillations or bumps, whereas for $Pr=7$ it does not. The difference in the behaviour of the rms velocities distinguishes the bursting and non-bursting regimes of the system. 

\begin{figure*}
\captionsetup[subfigure]{skip=5pt,font=normalsize,labelformat=empty}
\includegraphics[width=8cm]{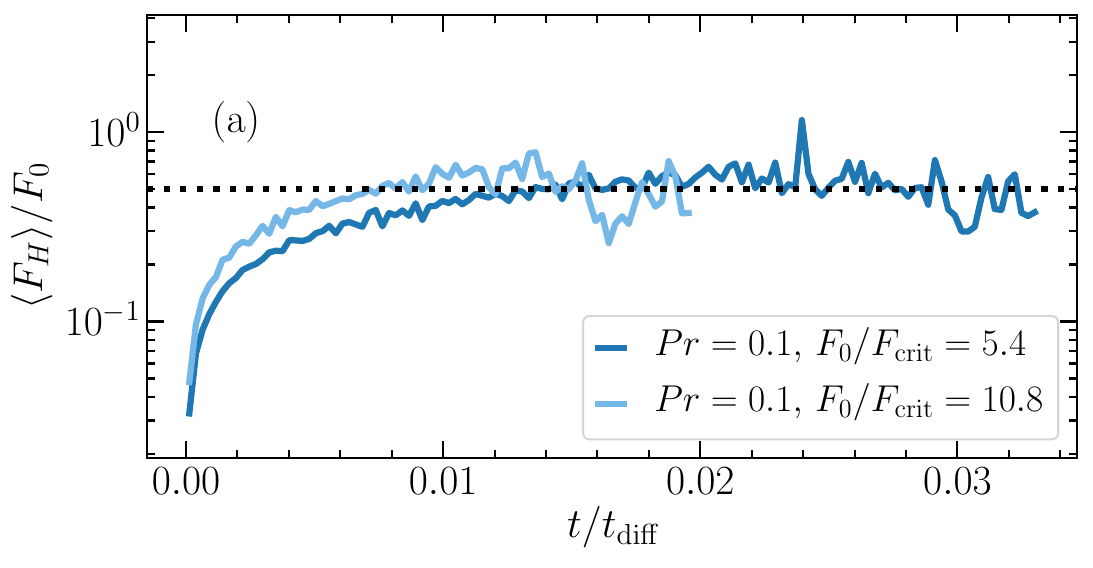}\hspace{0.01cm}
\includegraphics[width=8cm]{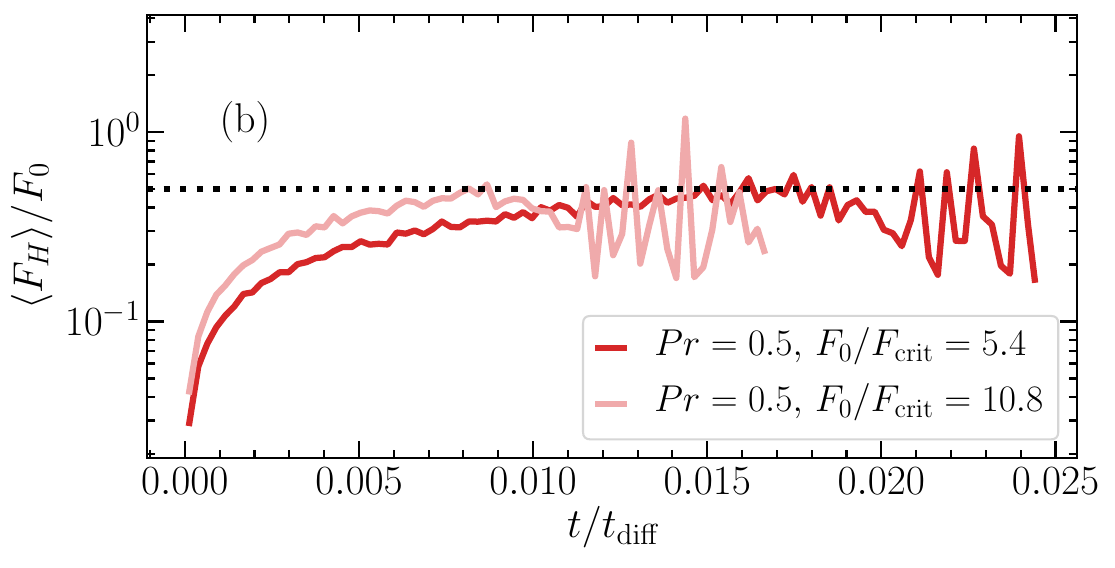}\\
\includegraphics[width=8cm]{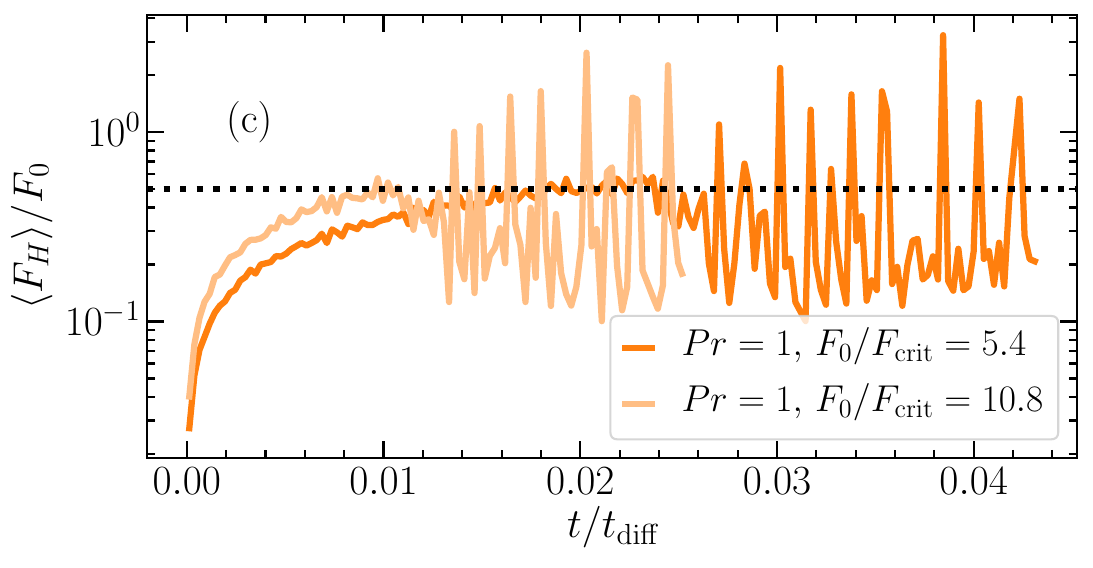}\hspace{0.01cm}
\includegraphics[width=8cm]{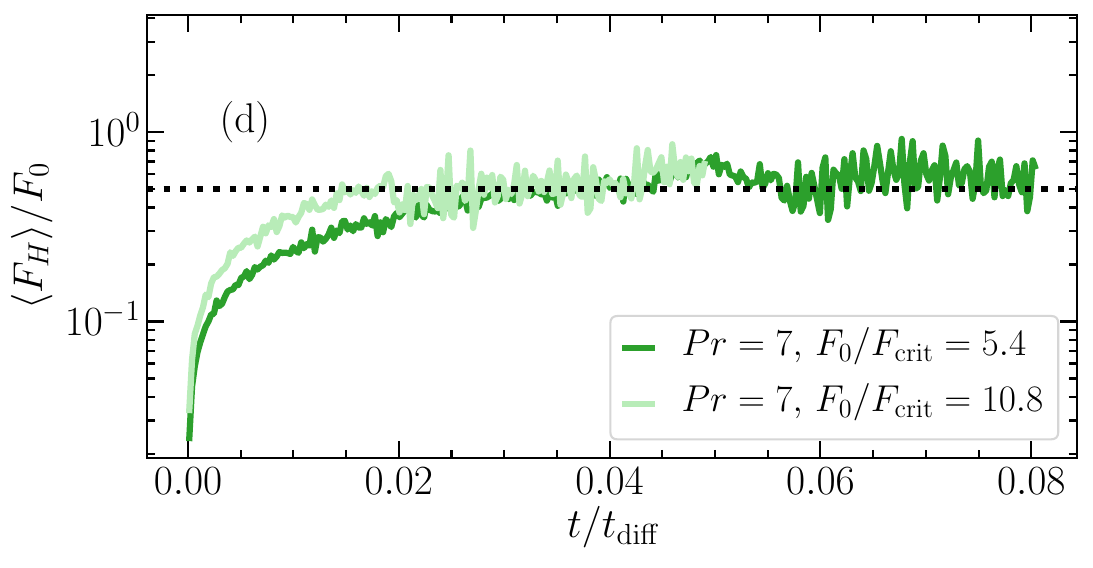}
\caption{Time series of the averaged heat flux (divided by the imposed flux at the top), $\langle F_H \rangle/F_0$ for $Pr = 0.1,\, 0.5,\, 1$, and 7 (panels a, b, c,  and d, respectively). Colors distinguish between $F_0/F_{\mathrm{crit}} = 5.4$ (dark) and $F_0/F_{\mathrm{ crit}} = 10.8$ (light). Note that panels do not share the same scale in the time axis. In all panels the dashed lines corresponds to $\langle F_H \rangle/F_0 = 0.5$, i.e., the expected averaged ratio if the whole fluid cools at a constant rate.} \label{fig_fh_time}
\end{figure*}

The bursting and non-bursting regimes have substantial differences in the vertical transport.  These differences are more clear when looking into the time series of the heat flux averaged over the whole domain
\begin{equation}
\langle F_H \rangle =  \rho_0 c_P\langle wT \rangle -  k\langle dT/dz \rangle\, ,\label{eq_Fh}
\end{equation}
where the first and second terms correspond to the flux carried by convective motions and diffusion, respectively. Figure \ref{fig_fh_time} shows time series of the ratio $\langle F_H \rangle/F_0$ for all our simulations. We observe for all the cases that as the convection zone grows, the convective contribution to the heat flux increases with time and dominates the magnitude of  $\langle F_H \rangle/F_0$. Once the zonal flow arises and becomes strong enough to disperse convective plumes and reduce the vertical kinetic energy, $\langle F_H \rangle/F_0$ decays. The subsequent evolution of the heat flux is different depending on $Pr$. For the cases $Pr \leq 1$, the vertical transport occur through discrete bursts whose intensity and frequency increase with $Pr$. Each burst is separated by a quiescent phase in which $\langle F_H \rangle/F_0 \approx 15-20\%$ of its value before the onset of the zonal flow. 

\begin{figure*}
\hspace{0.7cm}\includegraphics[width=6cm]{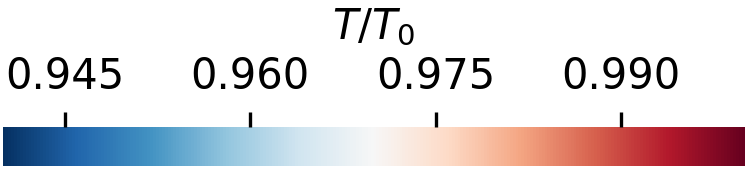}\\
\includegraphics[width=\textwidth]{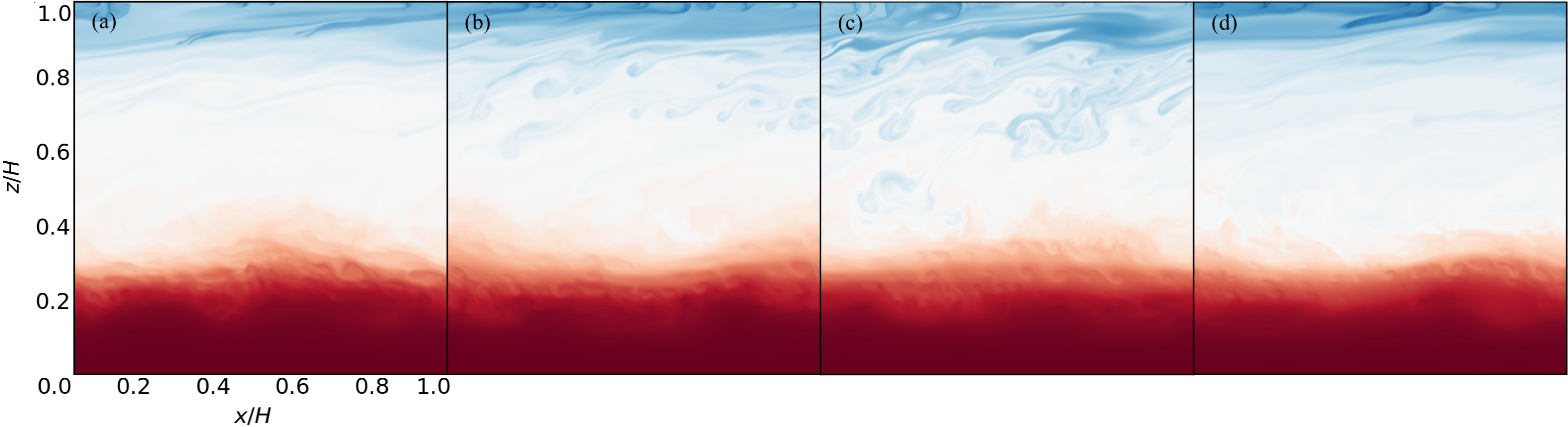}
\caption{Instantaneous snapshots of the temperature field (normalized to the initial temperature) during a single burst. Results are shown for the run using $Pr=0.5$ and $F_0/F_{\rm crit} = 10.8$. All panels share the same color scale.}
\label{fig_burst}
\end{figure*}

Figure \ref{fig_burst} shows the different stages of a burst. In the quiescent phase (a), convective plumes are constantly dispersed by the zonal flow. This flow suppresses convective instabilities until it has decayed sufficiently for convective plumes to appear again (b) and the fluid suddenly overturns (c). The fluid is energized by buoyancy and circulation motions, but once again, kinetic energy is transferred to the zonal flow, convective plumes get dispersed, and a new quiescent phase begins (d).

\begin{figure*}
\centering
\includegraphics[width=5.4cm]{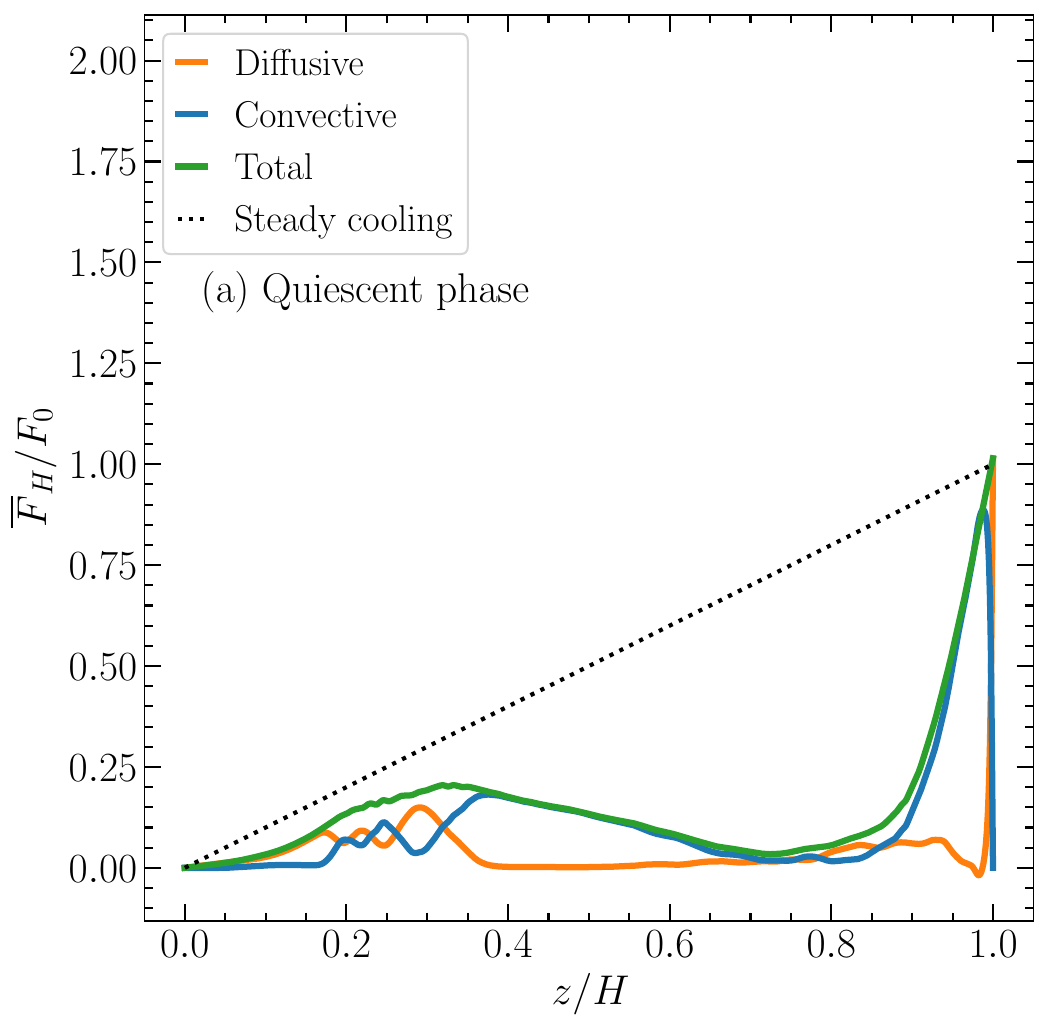}\hspace{0.005cm}
\includegraphics[width=5.4cm]{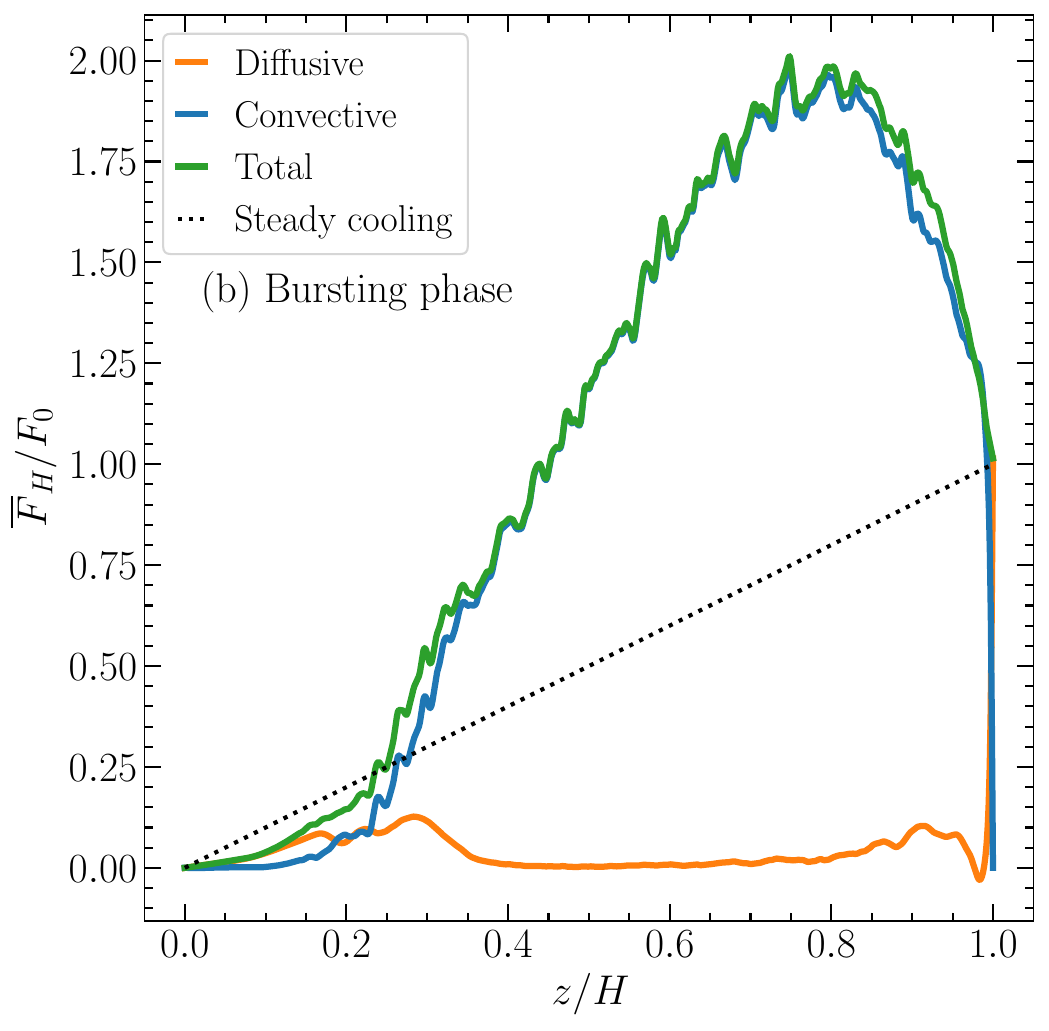}\hspace{0.005cm}
\includegraphics[width=5.4cm]{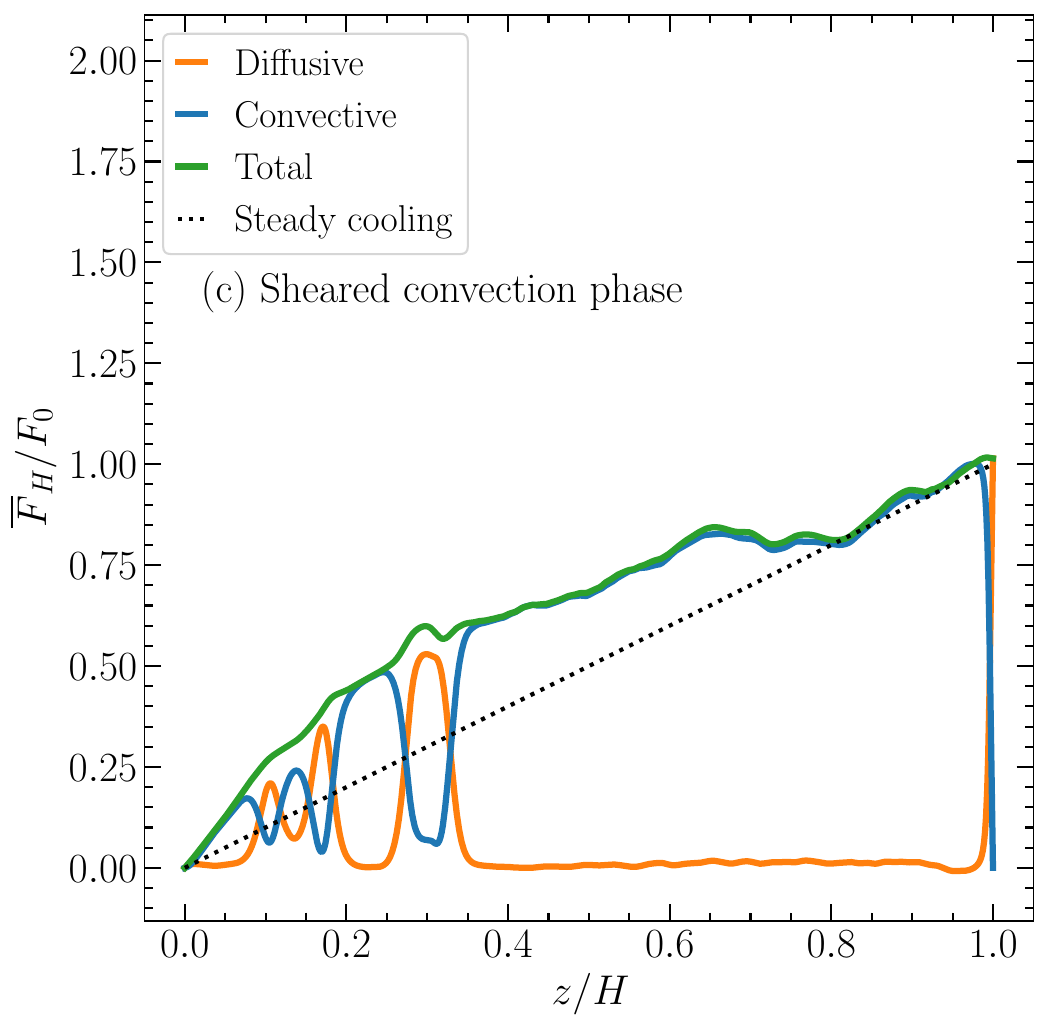}
\caption{Horizontally-averaged profiles of the ratio between the heat flux and the imposed flux, $\overline F_H/F_0$. Panel (a) shows the profiles during the quiescent phase, whereas Panel (b) shows the profiles for a particular burst during the active phases. Both panels show results for a typical burst in the simulation $Pr=1$, $F_0/F_{\rm crit} = 10.8$.  Panel (c) shows the profiles during the sheared convection phase observed in all simulations at $Pr=7$. The results are shown for the case $Pr=7$, $F_0/F_{\rm crit} = 5.4$ at $t/t_{\rm diff}=0.065$. In all panels, the green, blue, and orange lines correspond to the total, advective, and diffusive contribution to the flux, respectively. The dotted line corresponds to the expected profile if the whole fluid cools down at a constant rate, i.e., $\overline F_H/F_0 = z/H$.}\label{fig_f_profiles}
\end{figure*}

\begin{figure*}
\centering
{\includegraphics[width=7cm]{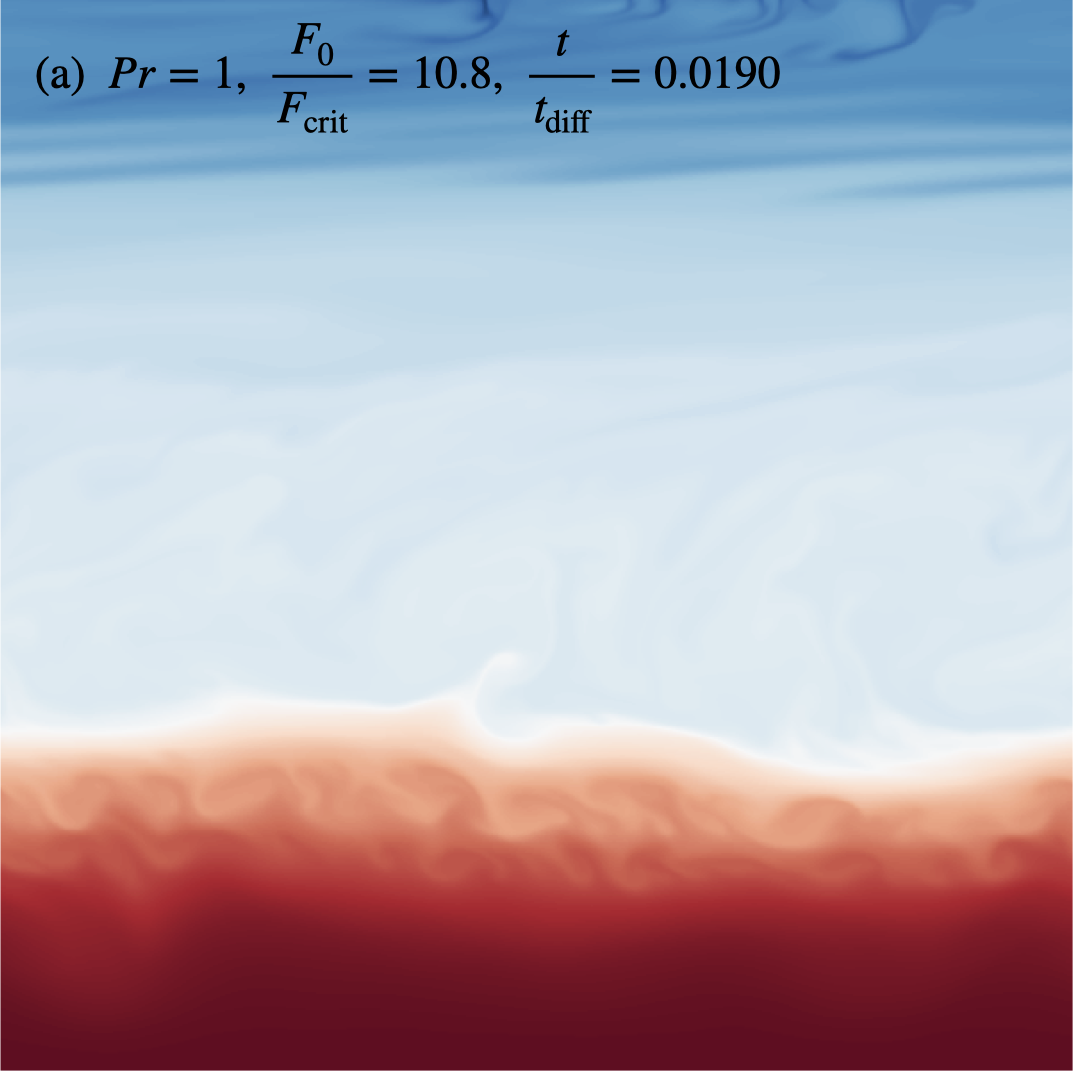}}
\hspace{0.01cm}
{\includegraphics[width=7cm]{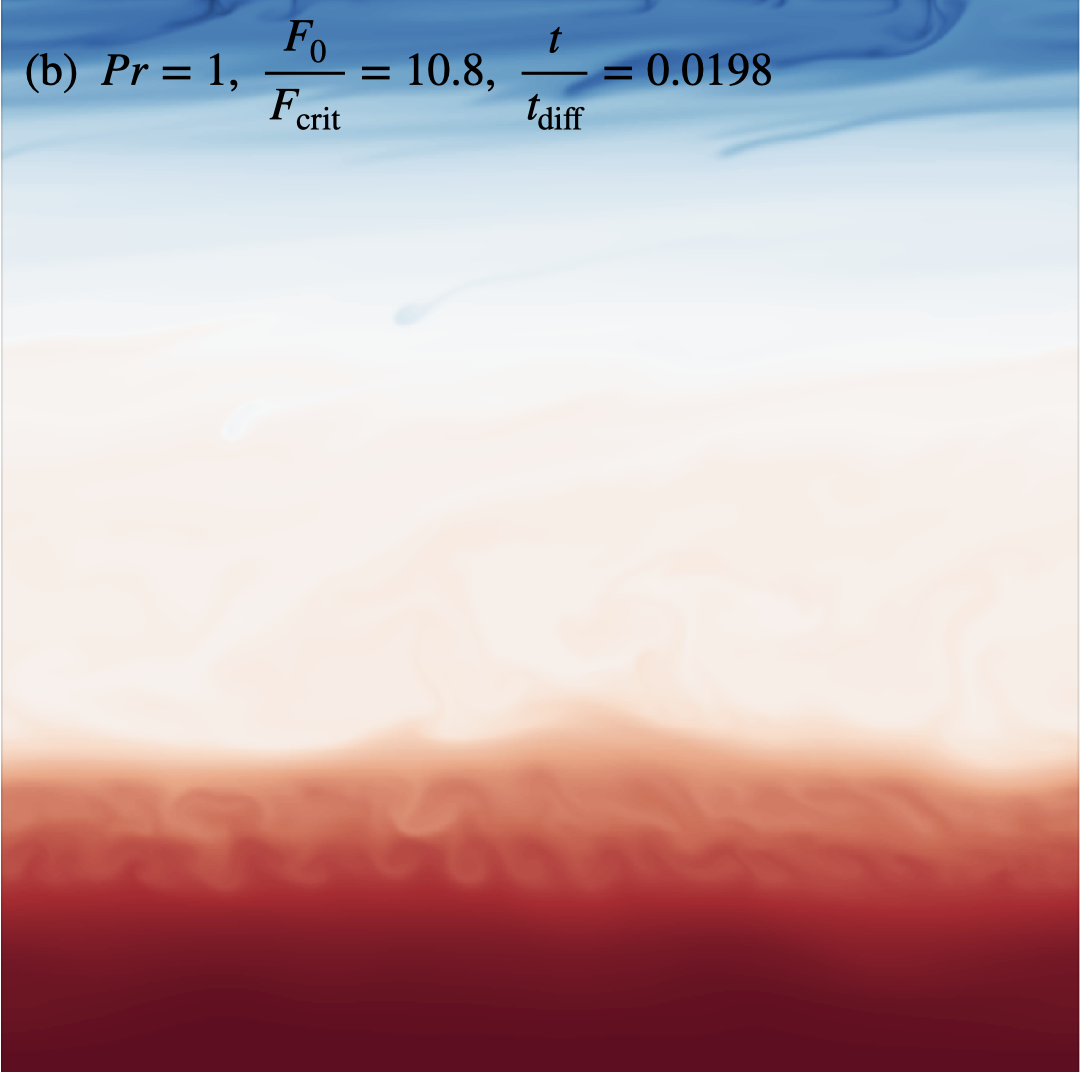}}\\
{\includegraphics[width=7cm]{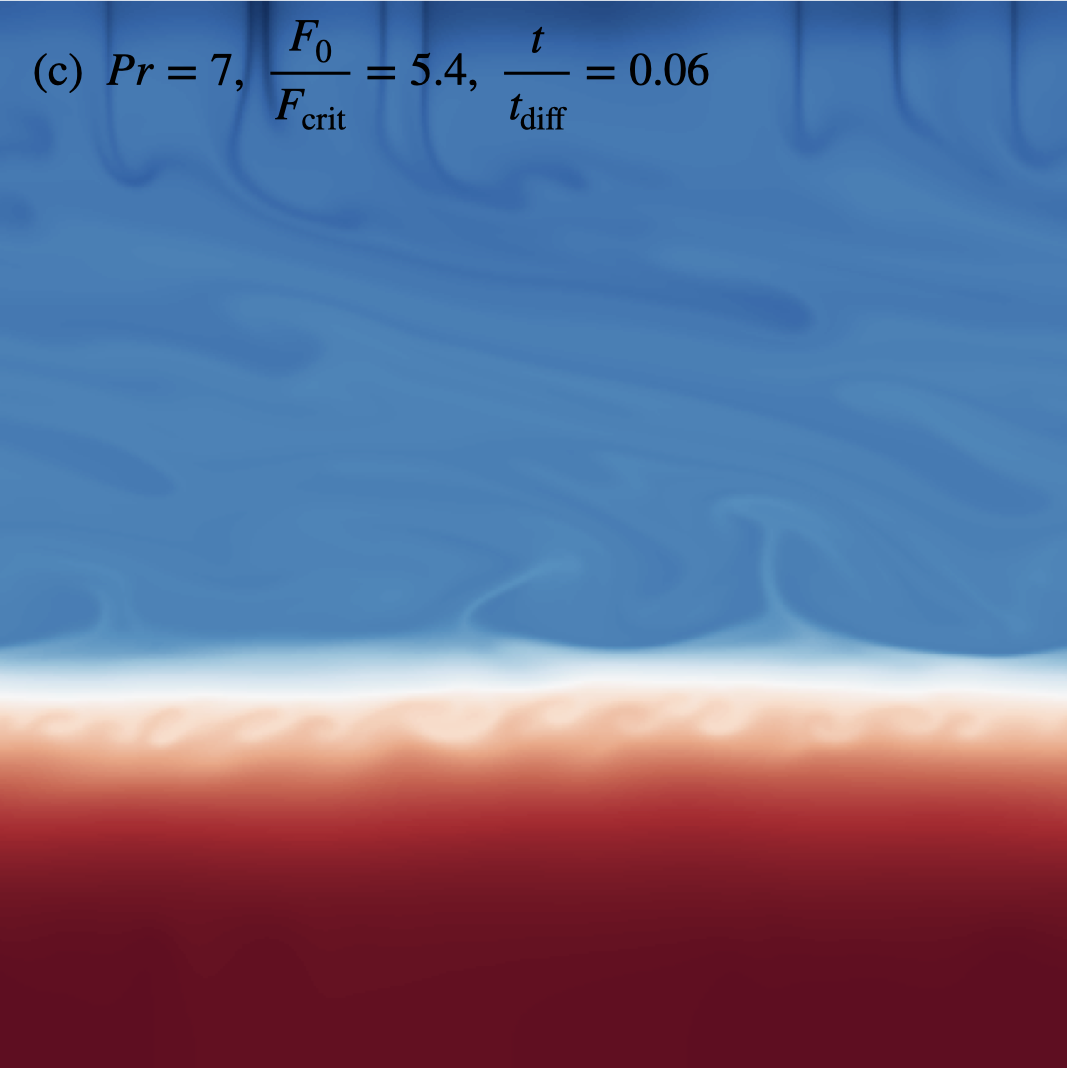}}
\hspace{0.01cm}
{\includegraphics[width=7cm]{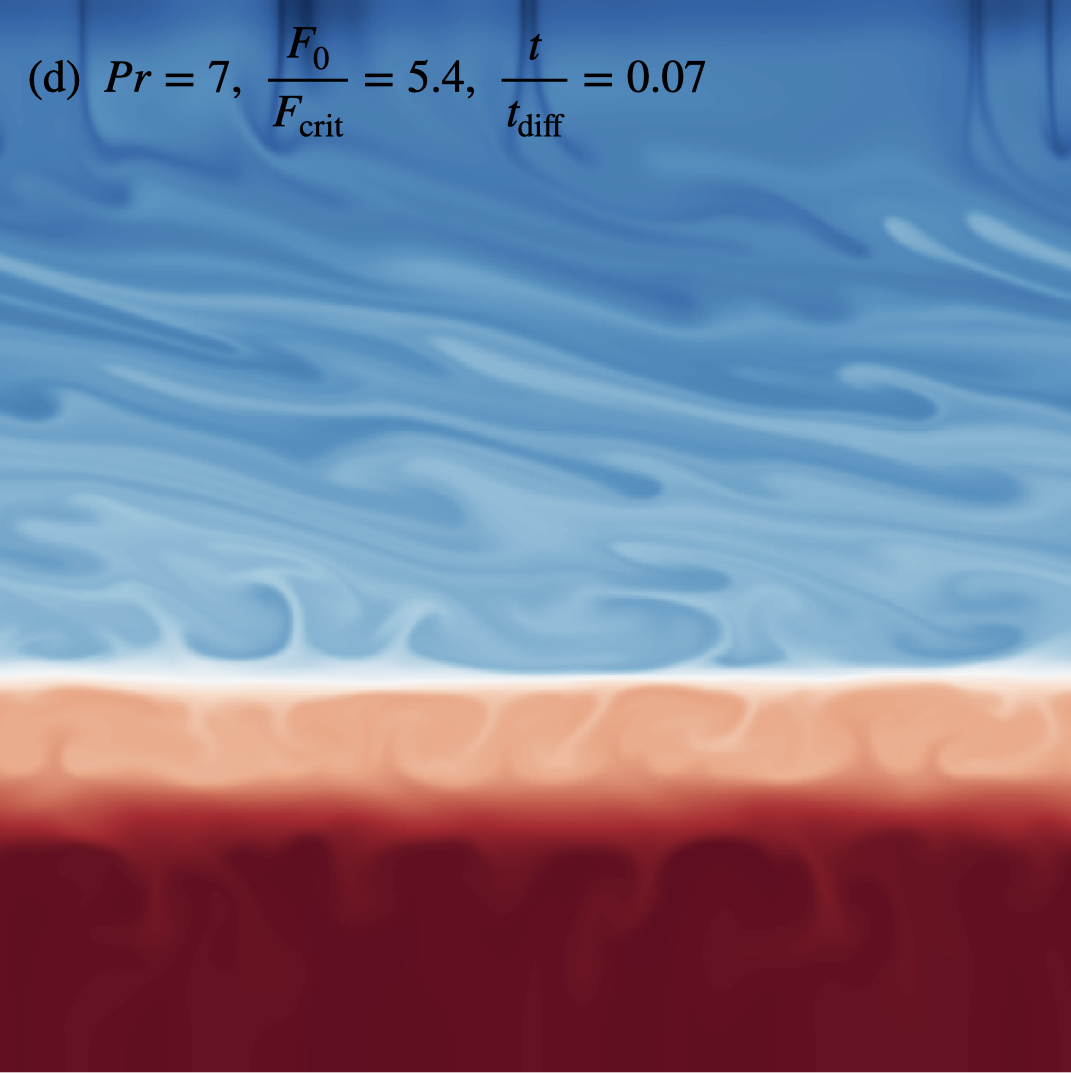}}
\caption{Instantaneous snapshots of the temperature field for the cases $Pr=1$, $F_0/F_{\rm crit} = 10.8$ (panels a and b), and $Pr=7$, $F_0/F_{\rm crit} = 5.4$ (panels c and d). The color scale is not the same in all panels. However, in each panel the darkest blue and red colors (the extremes of the color scale) represent the coldest and hottest fluid in the whole box at the given time. The times were chosen to show that secondary layers form and contribute the the averaged heat transport (in the quiescent phase for the case of panels a and b, and during the non-bursting regime for the case of panels c and d)}.
\label{fig_layers}
\end{figure*}

Note that although $\langle F_H \rangle/F_0$ decays during the quiescent phase, its value is not negligible since the averaged heat flux across the box is more than 10$\%$ of the imposed heat flux in all the cases. The reason for this can be explained using the flux profiles during the quiescent phase (see Fig. \ref{fig_f_profiles}a). We observe that near the top boundary (top of the convection zone) there are still convective motions that contribute to heat transport. This is expected since the fluid surrounding the top boundary is constantly cooling down by the imposed flux, thereby it has a permanent energy source to undergo convection. We also observe a smaller contribution to heat transport due to convective motions at the bottom of the outer convective layer ($z/H \approx 0.4$) and below it due to secondary convective layers at $z/H \approx 0.25$ (Fig. \ref{fig_layers}a-b). We recall that the flux time series in Fig. \ref{fig_fh_time} take into account the flux averaged over the whole box, however, even if we just consider the average over the outer convective layer, $\langle F_H\rangle/F_0$ would be still non-negligible. On the contrary, during the bursting phase as soon as the zonal flow weakens, the much colder fluid at the top sinks catastrophically in the way of a Rayleigh-Taylor plume, increasing significantly the heat flux (Fig. \ref{fig_f_profiles}b).

Finally, the cases $Pr=7$ are different. We do not observe bursts, and rather than cellular motions, vertically-sheared convective plumes dominate the flow within the convective layer. In those cases, the contribution of the convective flux to the total flux is significant at all times (Fig. \ref{fig_f_profiles}c). Further, as for the cases $Pr\leq 1$, the diffusive and convective flux profiles in Fig. \ref{fig_f_profiles}c (and also the snapshots in Fig. \ref{fig_layers}a-b) show that secondary convective layers form in the fluid, being responsible for the subsequent increase in $\langle F_H \rangle/F_0$. Since this work is focused on zonal flows and their properties, an analysis of layer formation and its evolution is going to be presented in a future paper. However, shear flows can destabilize fluids to Kelvin-Helmholtz and double diffusive instabilities, causing strong mixing and eventually forming multiple convective layers \citep[this mechanism is known as thermohaline shear instability, see, e.g., ][]{radko_2016,garaud_2017}. Although we did not test whether this mechanism is acting in our experiments, we observe Kelvin-Helmholtz billows near the convective boundary (Fig. \ref{fig_layers}c) prior to the formation of a second convection zone.

\subsection{Suppression of zonal flows in experiments with large aspect ratio} \label{sect_wider}

\begin{figure*}
\centering
\includegraphics[width=8.2cm]{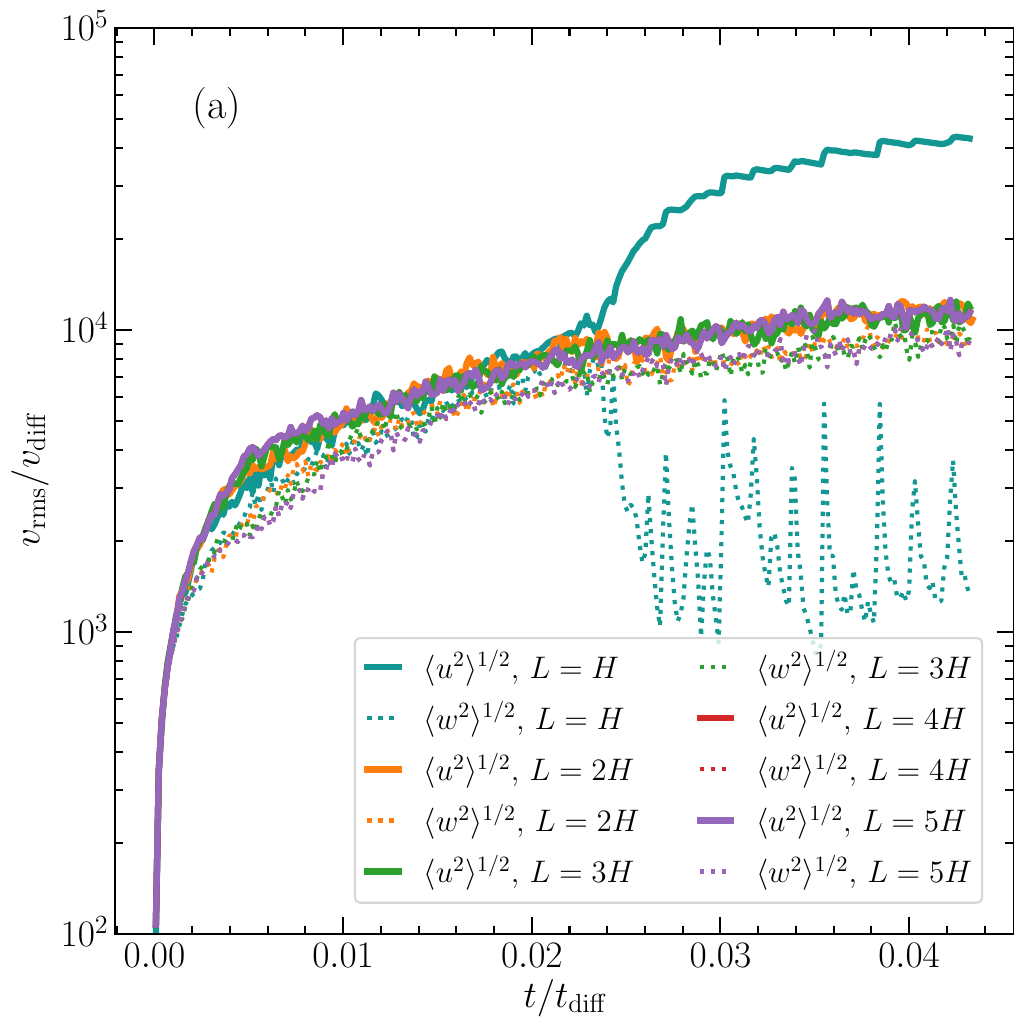}\hspace{0.01cm}
\includegraphics[width=7.95cm]{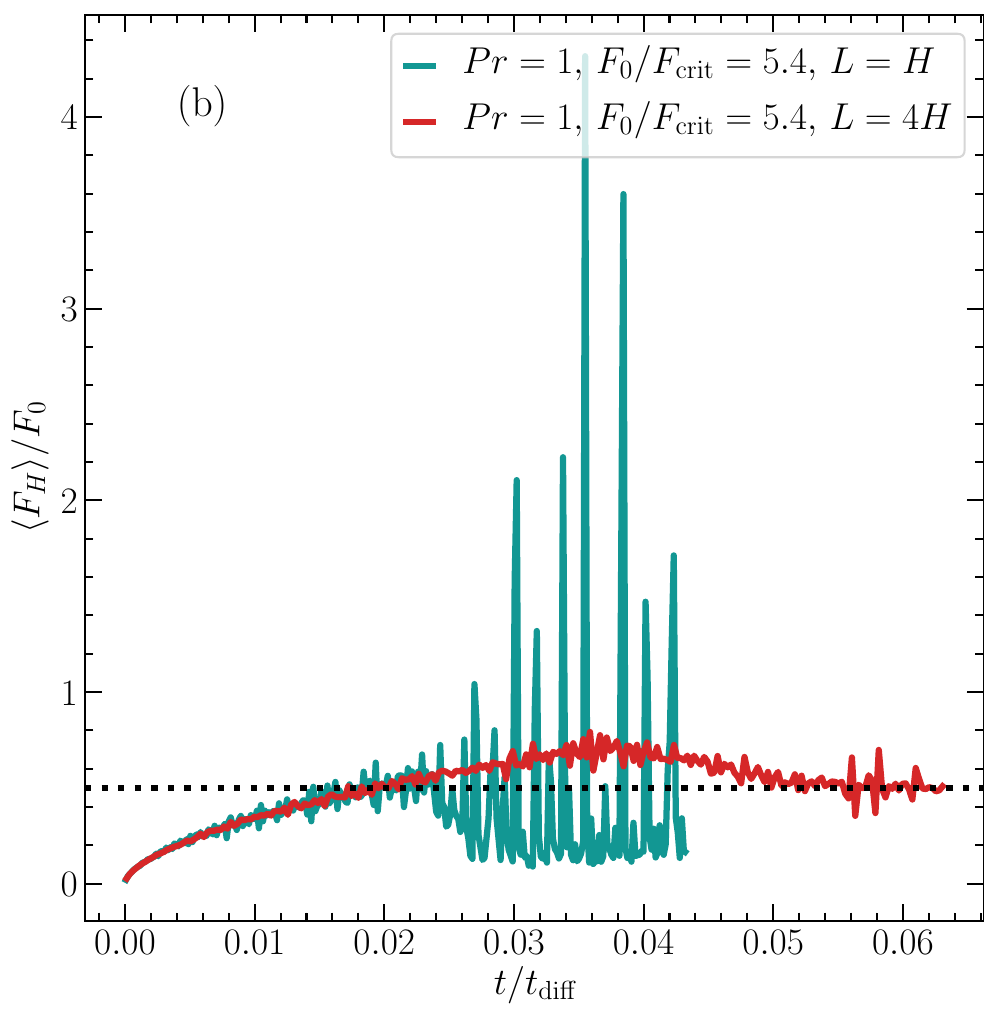}
\caption{Panel (a): Temporal evolution of the horizontal and vertical rms velocities ($\langle u^2\rangle^{1/2}$ and $\langle w^2\rangle^{1/2}$, using solid and dotted lines, respectively), for different domain width $L$ (using different colors). Panel (b): Time series of the averaged heat flux divided by the imposed flux at the top ($\langle F_H \rangle/F_0$), for simulations using different domain widths ($L=H$ and $L=4H$, as shown in the legends). In both panels, the results correspond to experiments using $Pr = 1$ and $F_0/F_{\rm crit} = 5.4$.} \label{fig_shear_supress}
\end{figure*}

We have shown that for computational domains with aspect ratio of one (i.e., $L/H = 1$), the zonal flow always appears when the aspect ratio of the convective layer, $L/h$, is smaller than approximately two (Fig. \ref{fig_aspect_ratio_Ra}a). For illustrative purposes, we show in Fig. \ref{fig_shear_supress}(a) the effect of increasing the width of the domain on the evolution of the rms velocities for runs using $Pr=1$ and $F_0/F_{\rm{crit}} = 5.4$. We find that for $L/H \geq 2$, both the horizontal and vertical rms velocities increase gradually in time unlike the experiments where $L/H=1$. Further, the curves lie on top of each no matter the value of $L$. Fig. \ref{fig_shear_supress}(b) shows the effect of increasing the width of the domain on the time series of the averaged heat transport. We find that the bursting regime disappears at large aspect ratio and the system evolves toward a state in which the whole fluid cools at a constant rate with $\langle F_H \rangle/F_0 \approx 0.5$.  We perform additional simulations for $Pr=0.1$, 1, and 7 and find the same behaviour. During the whole evolution of the simulations, the growing convective layer has an aspect ratio that is always much larger than two and thereby zonal flows are not expected to arise.

The suppression of the zonal flow has important consequences for the evolution of the flow. First, the vertical transport is always significant and never through quasi-periodic bursts. Second, the spatial structure of the flow is different. Whereas the simulations with aspect ratio of one have flows dominated by sheared convective plumes and bursts, the wider domain simulations exhibit convective cells which persist in time. Third, the convective layer never stops growing and reaches the bottom of the box, mixing the whole fluid.

\section{Summary and Discussion} \label{sec_conclusions}

We studied the onset and evolution of zonal flows when a convective layer propagates into a fluid with a stable composition gradient. We considered different values of the Prandtl number, $Pr = 0.1$, 0.5, 1, and 7. Our goal was to provide a novel way to study zonal flows and shear effects since the growing convective layer allows exploration of a wide range of values of the Rayleigh number and aspect ratios. Our results confirm and extend to convection with stable composition gradients at low $Pr$ previous findings in experiments of thermal and fingering convection.

In summary:

\begin{enumerate}

\item In simulations where the computational domain has an aspect ratio of one ($L/H=1$), zonal flows always arise, developing when the aspect ratio of the convective layer is smaller than approximately two. The critical aspect ratio for the onset of zonal flows seems to depend on $Pr$, being smaller at low $Pr$ (Fig \ref{fig_aspect_ratio_Ra}a).

\item Zonal flows are sustained by Reynolds stresses associated with tilted convective plumes. We find that the maximum magnitude of the stresses is limited by the vertical kinetic energy of the convective motions (Fig. \ref{fig_kinetic_energy}a). This result supports the hypothesis that energy is transferred from convective motions to the zonal flow.

\item Once the zonal flow sets in, it disperses convective plumes, reducing the vertical transport. As consequence, the convective layer stops growing. The stalling of the convection zone results in a saturation of the Rayleigh number toward a constant value in time, $Ra \sim 10^8-5\times 10^{9}$, where the smallest and largest values correspond to the cases using $Pr=7$, and 0.1, respectively (Fig. \ref{fig_aspect_ratio_Ra}b). 

\item As found in previous numerical simulations of pure thermal convection, the morphology and evolution of the flow depends on the Prandtl number. On the one hand, for $Pr\leq 1$ the flow organizes into discrete bursts in which convective plumes suddenly overturn quasi-periodically, with smaller transport between bursts (Figs. \ref{fig_fh_time}a-c, \ref{fig_burst}, and \ref{fig_f_profiles}a). On the other hand, for $Pr=7$ the flow consists of sheared convective plumes instead of bursts, and the convective transport is sustained at all times (Figs. \ref{fig_fh_time}d and \ref{fig_f_profiles}c).

\item The bursting regime is stronger at $Pr=1$ and weakens for $Pr < 1$ (Fig. \ref{fig_f_profiles}).

\item We observe the formation of secondary convective layers in all the simulations considered in this work (Fig. \ref{fig_layers}). The new layers contribute to the averaged heat transport at all times for both the non-bursting and bursting regimes (Fig \ref{fig_f_profiles}).

\item For wider domains ($L/H \geq 2$), the aspect ratio of the convective layer is always much larger than two and zonal flows never develop during the evolution of the simulations (Figs. \ref{fig_shear_supress}a-b). The absence of large scale horizontal flow means that the the growth of the convective layer is uninterrupted, and the fluid fully mixes in all the cases considered.

\end{enumerate}

We have shown that zonal flows arise in the time-dependent problem of a convective layer propagating into a stable fluid. Our work differs from previous studies by considering a stable composition gradient and the fact that convection is driven by a constant heat flux at the top boundary rather than a constant temperature contrast across the fluid depth.

We find the onset of the zonal flow when the aspect ratio of the convection zone is smaller than two and $Ra \sim 10^8$ - $10^{10}$. These values of $Ra$ are narrower than in previous work. For example, \citet{goluskin_johnston_flierl_spiegel_2014} found zonal flows can arise when $Ra \sim 10^4$ - $10^{10}$. A possible explanation could be the fact that in our problem convection occurs just in a portion of the fluid domain and the bottom of the convection zone is not stress-free, meaning that horizontal fluid motions at the bottom of the layer are decreased due to the interaction with the motion-less (stable) fluid below. On the contrary, in \citet{goluskin_johnston_flierl_spiegel_2014} convection occurs in the whole box, and the stress-free boundaries enhance at all times the horizontal fluid motions, which is favourable for instabilities that give rise to the zonal flow.

Despite the differences mentioned above, the zonal flow and its effects are similar to those reported in previous work. In particular, the transition from the bursting to the non-bursting regime seems to occur at $Pr\simeq 1$ no matter how convection is driven in the system of whether the fluid has composition gradients. We did not explore in detail the range of $Pr$ in which the system reaches the bursting regime. However, we find the intensity and frequency of the bursts decreases for $Pr < 1$.

Recent work by \citet{2020JFM...905A..21W} explored in detail the influence of the aspect ratio of the domain on the evolution of the zonal flow in thermal convection. By imposing initial conditions consistent with a linear shear-flow, and pure convective rolls, \citet{2020JFM...905A..21W} found that the zonal flow only persists or arises when the aspect ratio of the domain is smaller than a certain value depending on $Ra$ and $Pr$. For larger values, simulations initialized with convective cells do not develop sheared-flows, and the ones initialized with zonal flows transition to convective cells. Those results support previous findings by \citet{2014PhFl...26e4104F}, who demonstrated that in fluid domains of large aspect ratio, the tilting instability that enhances zonal flows can be suppressed due to the non-linear interaction of horizontal modes of the velocity field. Our simulations including composition gradients exhibit the same behaviour. We find that zonal flows only appear when the aspect ratio of the convective zone is smaller than two, and that it can be suppressed using wider fluid domains (see Figs. \ref{fig_aspect_ratio_Ra} and \ref{fig_shear_supress}).

The zonal flow and its effects are a problem for two-dimensional studies of convective transport and mixing in fluids where strong shear flows are not expected to appear. For example, layer formation and transport across diffusive interfaces in double-diffusive convection \citep{2015ApJ...815...42G,2020JFM...905A..21W}.  However, the fact that zonal flows do not appear at large aspect ratio suggests that two-dimensional simulations could still be useful to study convection and related problems involving stable composition gradients (such as convective overshoot). We did not explore different boundary conditions, however, additional simulations with the numerical set-up of \citet{2015ApJ...815...42G} but using a much larger aspect ratio would be of great interest in order to see if zonal flows can also be avoided in that situation.

\begin{acknowledgements}
We thank the two anonymous reviewers whose comments and suggestions helped improve and clarify this manuscript. This work was supported by  an NSERC Discovery Grant. J.R.F. acknowledges support from a McGill Space Institute (MSI) Fellowship. A. C. and J. R. F. are members of the Centre de Recherche en Astrophysique du Québec (CRAQ) and the Institut de recherche sur les exoplanètes (iREx). This research was enabled in part by support provided by Calcul Québec (calculquebec.ca), and Compute Canada (www.computecanada.ca). Computations were performed on Graham and Béluga.
\end{acknowledgements}

\appendix

\section{Resolution study} \label{sect_res_study}

We performed a convergence study in order to find the optimal resolution for all our simulations. Specifically, we compare the evolution of the thickness of the upper convection zone using $128^2$, $256^2$, $512^2$, and $1024^2$ modes (or $192^2$, $384^2$, $768^2$, and $1536^2$ grid points, respectively), for the case $Pr=1$ and $F_0/F_{\rm crit} = 10.8$. 

\begin{figure}
\includegraphics[width=8cm]{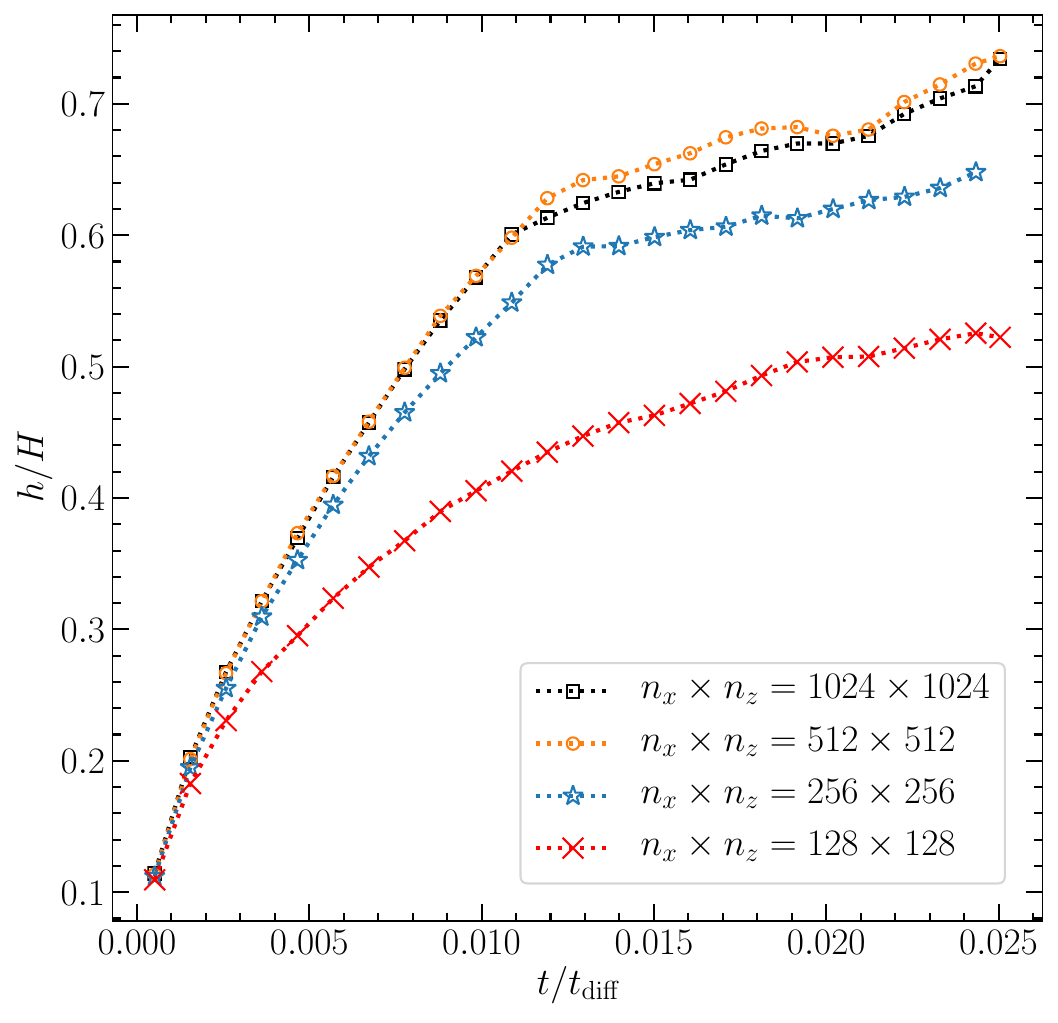}
\caption{Thickness of the upper convective layer (normalized to the height of the domain) as a function of
time (normalized to the thermal diffusion time) for different resolutions (number of modes) as shown in the labels. The parameters
used in this resolution study is $Pr = 1\, , \tau = 0.1$, and $F_0/F_{\rm crit} = 10.8$.}
\label{fig_res}
\end{figure}

Fig. \ref{fig_res} shows that curves using $512^2$ and $1024^2$ superpose perfectly (with minor differences of at most $0.7\%$ until $t/t_{\rm diff} \approx 0.01$, suggesting that results converge when using at least 512 modes (768 grid points) in each direction. The differences between the curves with 512 and 1024 modes observed for $t/t_{\rm diff}>0.01$ are due to either the random behavior of the turbulence once the zonal flow sets in and small eddies that are not resolved correctly with 512 modes.

As a double check, in the following we estimate the thickness of the boundary layers. A balance between advection by the interior flow and diffusion across the separating interface gives
\begin{equation}
\delta_T \sim \kappa_T^{1/2}\left(\dfrac{H_{\mathrm{conv}}}{v_{\mathrm{conv}}}\right)^{1/2}\hspace{0.25cm}\, , \hspace{0.25cm}\, \delta_S \sim \sqrt{\tau}\delta_T\, , \label{eq_bl}
\end{equation}
where $H_{\mathrm{conv}}$ and $v_{\mathrm{conv}}$ are the characteristic size and velocity of the convection zone, respectively. From mixing-length theory, the convective velocity is given by
\begin{equation}
v_{\mathrm{conv}} \sim (H_{\mathrm{conv}} a_{\mathrm{conv}})^{1/2} \sim H_{\mathrm{conv}}^{1/2} \left(g\alpha \delta T\right)^{1/2}\, , \label{eq_vc}
\end{equation}
where $a_{\mathrm{conv}} = g \alpha \delta T$ is the acceleration due to thermal buoyancy effects ($\delta T$). Substituting Eq. \eqref{eq_vc} in \eqref{eq_bl}, and introducing conveniently the kinematic viscosity $\nu$, we obtain
\begin{equation}
\dfrac{\delta_T}{H_{\mathrm{conv}}} \sim \left(\dfrac{1}{Ra Pr}\right)^{1/4}\, ,
\end{equation}
where $Ra = \alpha g H_{\mathrm{conv}}^3 \delta T/\kappa_T \nu$ is the well known Rayleigh number. In the standard problem of thermal convection, the extent of the convection zone is the size of the fluid domain (i.e., $H_{\mathrm{conv}} = H$) and convection is driven by a fixed temperature contrast across the fluid depth (i.e., $\delta T = T_{\rm bottom} - T_{\rm top}$). However, in our setup convection is driven by the temperature contrast across
the thermal boundary layer due to the imposed heat flux at
the top boundary, and $H_{\mathrm{conv}}$ grows in time limited by the initial composition
gradient, being $H_{\mathrm{conv}}(t)\leq H$. Therefore $Ra$ depends on time in
our problem. From the simulations, the largest Rayleigh number for the run using $Pr=1$ and $F_0/F_{\rm crit} = 10.8$ is $Ra \approx 10^9$, as shown in Fig. \ref{fig_aspect_ratio_Ra}b. Using $H_{\mathrm{conv}} \approx H$ we obtain $\delta_T/H \approx 0.004$ and $\delta_S/H \approx$ 0.0016. If we use 512 modes (768 grid points) in each direction, we resolve $\delta_T$ and $\delta_S$ with 3 and 1.2 grid points respectively, whereas if we use 1024 modes (1536 grid points), we resolve them with 6.1 and 2.5 grid points, respectively. Since $Pr\geq \tau$ in all our simulations, we resolve the viscous boundary layer with more points than the solute boundary layer.  We note that the numbers here are just an estimation and we find that 1024 modes are enough to resolve most of the flow structures, finding good agreement with previous work and laboratory experiments of convection in salty water \citep[e.g., ][]{molemaker_dijkstra_1997,2019ThCFD..33..383Z,2020PhRvF...5l4501F}.  

\bibliography{references}
\end{document}